\documentstyle[12pt,aaspp4]{article}
\def\arcsec {$^{\prime \prime}$}

\def\etal   {{et~al.\/}}
\def\HI     {H~{\sc I}}
\def\HII    {H~{\sc {II}}}

\def\mo     {{$M_{\odot}$}}

\input{psfig}

\begin{document}

\title{Testing CNO Enrichment Scenarios in Metal-poor Galaxies
	with HST Spectroscopy\footnote{Based on observations 
with the NASA/ESA {\it Hubble
Space Telescope} (HST)  obtained at the Space Telescope Science
Institute, which is operated by the Association of Universities for
Research in Astronomy, Inc., under NASA contract NAS 5-26555}
}
\author{Henry A. Kobulnicky\footnote{Visiting Astronomer, 
  German-Spanish Astronomical Center, Calar Alto, operated jointly 
  by the Max-Plank-Institut f\"ur Astronomie and the Spanish 
  National Commission for Astronomy.}\footnote{Presently at UCO/Lick 
  Observatory,  University of California, Santa Cruz, CA 95064}
	and Evan D. Skillman$^{2}$}
\affil{University of Minnesota \\Department of Astronomy \\ 
116 Church St. SE \\ Minneapolis, Minnesota 55455  \\ 
Electronic Mail: chip@astro.spa.umn.edu, skillman@astro.spa.umn.edu}
\authoremail{chip@astro.spa.umn.edu}

\author{To appear in {\it The Astrophysical Journal}}

\vskip 1.cm

\begin{abstract} 

Using {\it Hubble Space Telescope} ultraviolet and ground--based
optical spectroscopy, we measure the C/O and N/O ratios of three
metal-poor galaxies with similar metallicity but differing N/O.  These
observations, in conjunction with photoionization models, indicate that
the C/O ratios estimated from C~III] $\lambda$1909 and [O~III]
$\lambda$5007 lines are consistent with those measured from C~II]
$\lambda$2326 and [O~II] $\lambda$3727 lines.  Although inherently more
uncertain than C$^{++}$/O$^{++}$ due to poorly-known ionization
correction factors, we develop the use of the C$^{+}$/O$^{+}$ ratio as
a reasonable substitute diagnostic of the carbon abundance in
\HII\ regions.  The C~II] $\lambda$2326 multiplet is typically quite
weak, but its proximity to the [O~II] $\lambda\lambda$3726,3729 lines
makes these transitions a potential tool for measuring the carbon
abundance in high--redshift emission--line objects.

The derived chemical properties are consistent with a statistically
significant correlation between N and C abundances in metal-poor
extragalactic \HII\ regions, in the sense that systems with the largest
N/O ratios also have the highest C/O ratios.  This result is unexpected
if the dispersion in N/O among galaxies of similar metallicity is
caused by localized, temporary chemical enrichments from massive stars.
The presence of a correlation suggests, instead, that the majority of N
and C production is coupled, as expected from chemical evolution models
where C is produced predominantly by low mass stars and N is produced
predominantly by intermediate mass stars.  The C/N
ratio is, then, fixed by the initial mass function.
Since the occurance of localized
chemical ``pollution''  in star-forming galaxies appears to be low, the
relative overabundance of N in some galaxies compared to others at
similar metallicity is most plausibly interpreted as an indicator of
the global, secular chemical enrichment history.  As such, the N/O and
perhaps the C/O ratios can be used as a ``clock'' to estimate the time
since the last major episode of star formation.

\end{abstract}


\section{Introduction: Localized Chemical Enrichments
and CNO Variations in Metal-poor Galaxies} 

The chemical compositions of stars and gas within galaxies contain a
record of the star formation (SF) and nucleosynthesis history which may
be observed long after the stars responsible for metal production have
ceased to exist.  Considerable effort has gone into measuring the
relative abundances of heavy elements in Galactic and extragalactic
\HII\ regions (see reviews by Peimbert 1993; Pagel 1994), principally
carbon, nitrogen, and oxygen, which are the most abundant species after
hydrogen and helium).  These abundance measurements are powerful tools
that can be used to constrain models of stellar and big bang
nucleosynthesis (Pagel \etal\ 1992; Skillman \& Kennicutt 1993; Olive
\& Steigman 1995 and references therein), model the time evolution of
stellar populations, and examine the extent of gas inflow and outflow
in galaxies (Tinsley 1974, 1976; Carigi \etal\ 1995; Pagel 1994 and
references therein).

There is evidence, however, that spectroscopic abundance determinations
in some starburst galaxies are altered by short-term, localized
abundance enhancements, the nucleosynthetic products of massive star
populations. Such ``polluted'' regions may not reflect global galactic
abundances and therefore may be unsuitable as probes of the chemical
and star formation history (Kunth \& Sargent 1986; Pagel \etal\ 1986).
Models based on theoretical nucleosynthesis yields predict that massive
starburst clusters are capable of producing large localized chemical
enrichments (Esteban \& Peimbert 1995).  The starburst galaxy NGC~5253
exhibits the most pronounced evidence for chemical pollution of any
low-metallicity galaxy in the form of a three-fold N enrichment in the
central 40 pc (Welch 1970; Walsh \& Roy 1989; Kobulnicky \etal\ 1997).
In addition to this striking case of localized N overabundance, there
exists a large dispersion in N/O among metal poor systems
(7.8$>$12+log(O/H)$>$8.4), consistent with genuine N abundance
variations at a given metallicity (e.g., Garnett 1990 but see Thuan,
Isotov, \& Lipovetsky 1995 for an opposing view).

One scenario that might explain such scatter in N/O invokes
``pollution'' from the N-rich winds of Wolf-Rayet stars within the
present starburst (Pagel \etal\ 1986).  Yet, deliberate spectroscopic
searches seldom find localized chemical enrichments in the vicinity of
young star clusters in metal-poor galaxies (Kobulnicky \& Skillman 1997
and references therein).  A second scenario for the dispersion in N/O
at a given metallicity involves differing contributions from primary
and secondary nitrogen, which essentially amounts to variations of the
initial mass function from galaxy to galaxy.  A third scenario capable
of explaining the N/O dispersion while maintaining a universal IMF
involves a time--variable N/O ratio resulting from the delayed return
of N to the interstellar medium compared with the return of O.  A
fourth scenario for producing  N/O variations at constant O/H requires
preferential oxygen loss from galaxies with high N/O and more effective
oxygen retention in galaxies with low N/O.

To investigate the origin of the N/O dispersion among galaxies of
comparable metallicity and test whether N/O and C/O ratios
might serve as diagnostics of the recent star formation activity, 
we undertook a program of {\it Hubble Space
Telescope} and ground-based spectroscopy to measure the C and N
abundances in a sample of galaxies with differing N/O but constant
metallicity (as indicated by O/H).  A sample of four galaxies with
bright \HII\ regions and 12+log(O/H)$\approx$8.1 was selected from the
literature.  Two galaxies, UM469 and Tololo 1304-386 have relatively
high N/O ratios (log N/O $\sim-$1.2) while the other two, NGC~4861 and
Tololo 1345-420, have N/O ratios among the lowest in this metallicity
range (log(N/O)$\approx-$1.6 to $-$1.5).    Our principle objective was
to measure the C abundance of each object using the UV C~III]
$\lambda$1909 and C~II] $\lambda$2363 lines.  Section~2 summarizes the
HST and ground-based observational program undertaken to measure the C,
N, and O abundances.  Section~3 recounts the analysis procedures and
the derived chemical properties of each object, including a discussion
of the utility of the C$^+$/O$^+$ ratio for measuring carbon
abundances.  Sections~4 and 5 outline the implications for the
production of C and N in metal-poor galaxies and for the evolution of
N/O and C/O ratios in systems of similar metallicity.

\section{Optical and UV Spectroscopy} 
\subsection{HST FOS Observations and Reductions} 

Three of four targets, T1345-420, UM469, and NGC 4861 were observed
with the Hubble Space Telescope Faint Object Spectrograph (FOS) during
Cycle 6.  One object, T1304-386, could not be observed before the
removal of the Faint Object Spectrograph in 1997 January.  One location
in each object was observed through the 0.{\arcsec}86 circular aperture
of the Faint Object Spectrograph.  Wavelength coverage using gratings
G190H, G270H, G400H, and G570H ranged from 1600 \AA\ through 6800 \AA\,
with typical spectral resolutions of R=1300.  Exposure times were
selected to yield a strong detection of C~III] $\lambda\lambda$1907,1909
and O~III] $\lambda\lambda$1661,1666 for an accurate C/O abundance ratio
measurement.  In actuality, only upper limits on O~III]
$\lambda\lambda$1661,1666 were attained, but C~III]
$\lambda\lambda$1907,1909 was detected in all objects.  Table~1
summarizes the list of objects, gratings, approximate spectral
resolutions, and exposure times used at each location.

 The spectra were reduced using the standard HST FOS pipeline
procedures.  Spectra from different gratings were combined and the
continuum levels showed good agreement in the overlapping regions
(within 6\%) indicating a robust flux calibration.  This uncertainty
has not been propagated into the formal error budget since it
represents only the lack of agreement at the {\it ends} of each
bandpass, and would become significant only for spectral lines near the
edge of one spectrum {\it relative to} spectral lines near the edge of
the other overlapping spectrum.  The resulting calibrated spectra (not
corrected for reddening) are shown in Figures 1---3, smoothed with a
5--pixel boxcar for display purposes.

\subsection{Calar Alto 3.5m Observations and Reductions}

Optical spectra of NGC 4861 and UM469 were obtained with the Twin
spectrograph on the Calar Alto 3.5m telescope on the evenings of 1995
July 31 and 1997 January 12 respectively.  In 1995, the Tektronics
1024$^2$ format CCDs with 24 $\mu$m pixels were employed as detectors
on both the red and blue arms of the spectrograph yielding nominal
spatial scales of 0\arcsec.89 pixel$^{-1}$.  Useful data were collected
over the wavelength range of 3560 -- 5280 \AA\ in the blue and 5640 --
9170 \AA\ in the red.  Gratings of 300 line mm$^{-1}$ in the blue and
270 line mm$^{-1}$ in the red resulted in dispersion scales of 3.44
\AA\ pixel$^{-1}$ in the blue and 3.88 \AA\ pixel$^{-1}$ in the red.
The FWHM of the seeing averaged 1.2-1.5\arcsec.  Observations were
obtained with a 2.1$^{\prime\prime}$ wide slit.

During the 1997 run, the detectors were 800 x 2048 CCDs with 15 $\mu$m
pixels yielding nominal spatial scales of 0\arcsec.56 pixel$^{-1}$.
Gratings of 300 line mm$^{-1}$ in the blue and 270 line mm$^{-1}$ in
the red resulted in dispersion scales of 2.16 \AA\ pixel$^{-1}$ in the
blue and 2.43 \AA\ pixel$^{-1}$ in the red.  Useful data were collected
over the wavelength range of 3400 -- 5400 \AA\ in the blue and 5700 --
9600 \AA\ in the red.  Observations were obtained with a
2.1$^{\prime\prime}$ wide slit and spectra were extracted using a
4-pixel (3.5\arcsec) wide aperture.  Due to mechanical problems with
the telescope focus, spatial resolution was not limited by seeing, and
it varied over the night, averaging $\sim$3\arcsec.

On both nights the sky appeared hazy but uniform.  Bias frames, dome
flats, twilight sky flats, and He-Ar arc lamp exposures were taken at
the beginning and end of the nights.  Each night five standard stars
from the list of Oke (1990) were observed with a slit width of
3\arcsec.6, which is the maximum allowed by the spectrograph set-up.
Standard reduction procedures were followed using the programs
available within the IRAF\footnote{IRAF is distributed by the National 
    Optical Astronomy Observatories,
    which are operated by the Association of Universities for Research
    in Astronomy, Inc., under cooperative agreement with the National
    Science Foundation.}
 software.  During the flux calibration stage,
an atmospheric extinction law was derived from the standard star data.
This was done self-consistently, i.e., the standard star data from both
the red spectra and blue spectra were used simultaneously to derive the
extinction law.  Observations were taken close to the parallactic angle
to avoid problems of differential atmospheric refraction (particularly
important given the long wavelength coverage of the observations).
Three 600 sec exposures of each object were obtained and averaged after
rejecting pixels affected by cosmic rays.  A series of 1-D spectra were
extracted for each object.  Since it was not possible to match the
ground-based apertures spatially to the HST aperture, we selected the
spectrum nearest the location of the HST FOS aperture with the best
signal to noise.  For NGC 4861 the extracted region measured
2.1\arcsec\ $\times$ 3.5\arcsec.  For UM469 the extracted region
measured 2.1\arcsec\ $\times$ 10.0\arcsec.  Figures~4---5 show the
resulting spectra for for NGC 4861 and UM469, uncorrected for
reddening.

\section{Analysis and Abundance Computation}
\subsection{Extinction Correction}

Reddening, electron temperatures, densities, ionic abundances and total
elemental abundances (except carbon and silicon) are derived from
optical emission lines as described in Kobulnicky \& Skillman (1996).
Briefly, corrections for reddening  and underlying stellar Balmer
absorption  are computed using the standard procedures.

\begin{equation}
{{I(\lambda)}\over{I(H\beta)} } = { {F(\lambda)}\over{F(H\beta)} }
10^{c(H\beta)f(\lambda)}  \ , 
\end{equation}

\noindent where $I$ is the true de--reddened flux at a given
wavelength, $F$ is the observed flux at each wavelength, c(H$\beta$) is
the logarithmic reddening factor, and $f(\lambda$) the reddening
function (Seaton 1979 as parameterized by Howarth 1983) appropriate to
the Milky Way.  Shortward of 3200 \AA, the standard Galactic extinction
law diverges from that found in the LMC and SMC which are the only
metal poor objects with measured reddening curves.  Since the fluxes of
the UV lines relative to H$\beta$ are sensitive to the adopted values
of c(H$\beta$) and the assumed UV extinction curve, each line of sight
deserves careful attention.  Tables~2 and 3 list $f(\lambda)$'s for the
SMC and Galactic reddening laws, and the dereddened line strengths for
the HST FOS and Calar Alto spectra corrected for reddening relative to
H$\beta$, according to the prescriptions outlines below.  Corrections
for 0 to 2 \AA\ of underlying stellar H (but not He) absorption have
been applied based on the theoretical-to-observed
ratios of higher order Balmer lines in each object's spectrum.

{\it NGC~4861}

The values of c(H$\beta$) derived from the HST FOS and Calar Alto
Balmer line ratios are 0.20$\pm$0.06 and 0.11$\pm$0.06 respectively.
The best estimate of the underlying stellar Balmer absorption is 0.0 \AA\
from the FOS and 0.0 \AA\ from the Calar Alto data.  Dinnerstein \& Shields
(1986) report c(H$\beta$)=0.34 for this object. Given the vastly
different aperture sizes for each spectrum, the lack of agreement is
not terribly surprising if the extinction is spatially variable.  In
the following analysis, we will use the HST result,
c(H$\beta$)=0.20$\pm$0.06.

Galactic reddening maps based on star counts and \HI\ column density
(Burstein \& Heiles 1984) suggest E(B-V)=0.0, and thus c(H$\beta$)=0.0
toward this line of sight, so that nearly all of the extinction must be
internal to NGC~4861.  Since NGC~4861 has metallicity
12+log(O/H)$\approx$8.0, much like the SMC, we adopt an SMC type
reddening law (Pr\'evot \etal\ 1984 as parameterized in Kobulnicky
\etal\ 1997) to deredden the UV lines.

{\it UM~469}

The values of c(H$\beta$) derived from the HST FOS and Calar Alto
Balmer line ratios are 0.32$\pm$0.10 and 0.46$\pm$0.06 respectively.
The best estimate of the underlying stellar Balmer absorption is 1.5 \AA\
from the FOS and 2.0 \AA\ from the Calar Alto data.  The reasonable
agreement despite the vastly different aperture sizes for each
instrument probably indicates that the \HII\ region in UM~469 is
unresolved due to its great distance ($z=0.057$).  That we find
comparable  H$\beta$ fluxes through both apertures (actually greater
through the FOS!) is consistent with this hypothesis.  These
c(H$\beta$) values also compare favorably to the
c($H\beta$)=0.36$\pm$0.18 cited by Campbell, Terlevich, \& Melnick
(1986, hereafter CTM).  Galactic reddening maps of Burstein \& Heiles
(1984) suggest E(B-V)=0.01 ( c(H$\beta$)$\approx$0.01) in this part of
the sky, so that the measured c(H$\beta$) must be due mostly to
extinction within UM~469.  Additional evidence for this conclusion
comes from examination of the Mg~II $\lambda$2800 doublet seen in
absorption at Galactic velocities near 2800 \AA\ and at the redshift of
UM~469 near 2960 \AA.  Since the equivalent widths of the redshifted
Mg~II lines are more than 4 times greater than those seen at Galactic
velocities, most of the intervening medium lies within UM~469.
Consequently, we adopt an SMC type reddening law to deredden the UV
lines.

{\it T1345-420}

The value of c(H$\beta$) derived from the HST FOS Balmer line ratios is
0.25$\pm$0.06.  For complementary ground-based spectroscopy, we adopt
the measurements of Campbell, Terlevich \& Melnick (1986) who find a
similar value, c(H$\beta$)=0.17$\pm$0.05.  The reddening maps of
Burstein \& Heiles (1984) suggest E(B-V)=0.085 in this direction,
corresponding to c(H$\beta$)=0.13.  Thus, about half of the extinction
should arise in the Galaxy and the other half within T1345-420.
Consequently, we adopt a reddening law which is 50\% SMC and 50\%
Galactic.

\subsection{Uncertainties and Error Propagation}

The uncertainties on the optical emission lines in Table~3 are
1$\sigma$ errors which include contributions from photon statistics,
flatfielding, flux calibration, detector noise, sky background, and
reddening as described in Kobulnicky \& Skillman (1996).  For the 
HST data in Table~2, the listed uncertainties are computed empirically 
from
the RMS in adjacent continuum portions of the spectrum 
and the width of the line.  The listed uncertainties also include
include a contribution due to reddening.  All line strength uncertainties
are propagated by Monte-Carlo techniques into uncertainties on the
physical conditions and chemical abundance ratios as described below.
The tabulated 1$\sigma$ uncertainties reflect the quantifiable
statistical uncertainties only.  They represent a realistic 
lower bound on the true uncertainties which may have additional
contributions from 
systematic effects such as temperature and density fluctuation within
the aperture (e.g.  Peimbert, Sarmiento, \& Fierro 1991).  However, in
longslit spectroscopic surveys covering $\sim$1 linear kpc at spatial
resolutions of $\sim$20-40 pc in the nearby irregular galaxies NGC 4214
(Kobulnicky \& Skillman 1996) and NGC~1569 (Kobulnicky \& Skillman
1997), variations in physical conditions and chemical composition are
well within the formal statistical uncertainties.  This suggests that,
if significant fluctuations in physical conditions are present in those
systems, they evidently are important only on scales much smaller than
the typical resolution element ($<$40 pc) or are smaller than the
statistical uncertainties.  Consequently, we quote only the statistical
uncertainties and urge discretion in the interpretation of errors.

\subsection{Physical Conditions and Abundance Derivation}

The derived physical conditions and ionic and total abundances are
compiled in Tables 4, 5, and 6 for NGC 4861, UM~469, and T1345-420.
The electron density lies in the low-density limit in all cases, as
computed from the [S~II] $\lambda\lambda$ 6717,6731 line ratios.  We
adopt the two--temperature two-zone approximation commonly used,
consisting of an inner, high ionization (O$^{++}$) zone and an outer
zone of lower ionization (O$^{+}$).  The electron temperature,
$T_e(O^{++})$, of the high ionization zone is calculated from the ratio
of [O~III] [$\lambda$4959 + $\lambda$5007]/$\lambda$4363, while the
temperature of the low ionization zone, $T(O^+)$, is found using a
relation between $T_e(O^+)$ and $T_e(O^{++})$ (Pagel \etal\ 1992;
Skillman \& Kennicutt 1993) based on photoionization models
(Stasi\'nska 1990; Skillman 1989). 

Optical line ratios and computed electron temperatures are used to
determine He, N, O, and S abundances as described in
Kobulnicky \& Skillman (1996) making use of updated atomic data
in the code detailed in Shaw \& Dufour (1995).  Neither the O~III]
$\lambda\lambda$1661,1666 multiplet nor the N~III] $\lambda\lambda$1750
multiplet are detected in any of the objects, and the upper limits do
not allow any useful constraints to be placed on the N$^{++}$/O$^{++}$
or C$^{++}$/O$^{++}$ ratios.  Consequently, N/O is determined from the
ratio of the optical [N~II] $\lambda$6584/[O~II] $\lambda$3727 lines or
the [N~II] $\lambda$6584/[O~II] $\lambda\lambda$7320,7330 lines where
Calar Alto spectra are available.  Since the spectral resolution of the
FOS is not sufficient to separate the weak [N~II] $\lambda$6584 from
H$\alpha$, the ground-based Calar Alto data are used to compute the N
abundances.  C/O ratios are computed from the C~III]
$\lambda\lambda$1907,1909 and [O~III] $\lambda$5007 line strengths using

\begin{equation}
{ {X(C)}\over{X(O)} } = { {X(C^{++})}\over{X(O^{++})} }\times ICF =
   { {I_{1909}}\over{I_{5007}}} { {\epsilon_{5007}}\over{\epsilon_{1909}} } \times ICF
   \simeq ICF \times {{I_{1909}}\over{I_{5007}}}  {0.059 } e^{ 4.659/t_4 } \ \ .
\end{equation}

\noindent $ICF$ is the ionization correction factor, $t_4$ is the electron
temperature in units of 10,000 K, and $\epsilon$ is the volume
emissivity of a given transition.  The constant 0.059 involves an
interpolation for the temperature--dependent ratio of collision
strengths, $\Omega$(1909) and $\Omega$(5007) (Dufton \etal\, 1978 as
tabulated in Pradhan \& Peng 1996).  This term varies from 0.62 to 0.57
over the range 12,000 K and 14,000 K.  The assumption that ${
{X(C)}\over{X(O)} } \simeq\ { {X(C^{++})}\over{X(O^{++})} }$ is
justified from photoionization models (Garnett \etal\ 1995a, Figure 2)
which show that when the ionic fraction of doubly ionized oxygen,
X($O^{++}$), is less than $\approx$0.90, this approximation is accurate
to better than 20\% (0.08 dex) so that the ICF is between 1.00 and
1.15.  Examination of X($O^{++}$) for each object in Tables~4---6 show
that this condition is satisfied in all cases. The $O^{++}$ fractions
are 0.85, 0.78, and 0.85 for NGC~4861, UM469, and T1345-420
respectively.  Given that the
ionization correction factor is uncertain, but near unity, 
we assume $C^{++}/O^{++}$ = $C/O$ and do
not apply an ICF (or equivalently, ICF=1.0) in the text and tables.

Although the dominant ionization state of C and O in \HII\ regions
is C$^{++}$ and O$^{++}$, the C~II] $\lambda$2326 multiplet and [O~II]
$\lambda\lambda$3727,3729 lines can serve as a valuable check on the C/O 
ratio using
\begin{equation}
{ {X(C)}\over{X(O)} } = { {X(C^{+})}\over{X(O^{+})} }\times ICF =
   { {I_{2326}}\over{I_{3727}}} { {\epsilon_{3727}}\over{\epsilon_{2326}}}
   \times ICF \simeq ICF\times {{I_{2326}}\over{I_{3727}}}  {0.49 } e^{ 2.32/t_4 } \ \  .
\end{equation}

\noindent The constant 0.49 involves a linear interpolation for the
temperature--dependent ratio of collision strengths, $\Omega$({2326})
 and $\Omega$({3727}) (Blum \& Pradhan 1992 as tabulated in Pradhan \&
Peng 1996).  This term varies from 0.51 to 0.48 over the range 11,000 K
and 14,000 K of interest here.

The line ratio ${I_{2326}}/{I_{3727}}$ can also be used to estimate the
C/O ratio when data on the higher ionization diagnostics, C~III]
$\lambda$1909 and O~III] $\lambda$1666, are not available.  Although the
C~II] $\lambda$2326 multiplet is typically quite weak, it is detected
in all three of our targets and it is, in some respects, better for
measuring  C/O since it is less sensitive to reddening and temperature
uncertainties than the ${I_{1909}}/{I_{5007}}$ line ratio.  For the
objects studied here, the largest quantifiable source of uncertainty on the
C$^+$/O$^+$ ratio is low signal-to-noise on the C~II] $\lambda$2326
line.  However, systematic effects due to ionization correction
uncertainties and line blending make additional
contributions to the error budget.  Even in the ideal case of very high
S/N spectra, ionization correction
uncertainties and line blending will limit the usefulness of this
approach.  

Since the C~II] $\lambda$2326 multiplet is blended with the [O~III]
$\lambda\lambda$2321,2331 lines at the resolution of the FOS, a
correction to the C~II] line strengths is necessary.  The contribution
of [O~III] $\lambda\lambda$2321,2331 to the C~II] $\lambda$2326 feature
in the FOS spectra is on the order of 20\%.  For each object we have
estimated fractional contribution of [O~III] $\lambda\lambda$2321,2331
using the temperature and density appropriate for each case, while
assuming X(C$^{+}$)=0.1, X(O$^{++}$)=0.9, and C/O=0.37.  For NGC~4861,
UM469, and T1345-420, the [O~III] lines contribute 22\%, 21\%, and 20\%
respectively, to the total blended feature.  Table~2 lists the
estimated line strengths for [O~III] $\lambda\lambda$2321,2331 and the
adjusted line strengths for C~II] $\lambda$2326.  Note that the quoted
uncertainties on these lines do not include the contribution from
errors introduced by assuming the values for X(C$^{+}$), X(O$^{++}$),
and C/O as listed above.  Because of the uncertainties due to line
blending, we present C/O ratios derived from C$^+$ and O$^+$ lines as a
valuable {\it secondary} means of abundance measurement only.  We adopt
the C/O ratios derived from C$^{++}$ and O$^{++}$ lines as the
best-estimate values in all later analyses.

The second potential systematic effects arises because C$^+$is a trace
species within highly-ionized \HII\ regions, but ubiquitous in warm,
photo-dissociation regions.  Ionization correction factors may
potentially be large, and in general, the  higher ionization
diagnostics, C~III] $\lambda$1909 and O~III] $\lambda$1666 should
always be used when possible.  To investigate the magnitude of the
ionization correction factor needed, we have compiled \HII\ region
photoionization model results from Stasi\'nska (1990) and Stasi\'nska
\& Leitherer (1996).  In Figure~6 we plot the ratio of the ionic
fractions, X(C$^+$)/X(O$^+$), versus the fraction of doubly ionized
oxygen, X(O$^{++}$).  [X(C$^+$)/X(O$^+$)]$^{-1}$ is the ionization
correction factor that converts the observed C$^+$/O$^+$ ratio into
C/O.  A wide range of model parameters are explored at a metallicity of
Z=0.2 Z$_\odot$, appropriate to the observed galaxies.  In the top
panel, lines connect models (Stasi\'nska 1990) with the same
temperature stars while symbols denote models with similar ionization
parameter, (corresponding roughly to the number of stars ionizing the
\HII\ region in her models).  X(C$^+$)/X(O$^+$) lies in the narrow
range 0.80 to 0.95 for 0.75$<$X(O$^{++}$)$<$0.85 which is the regime of
interest for the three targets.  Although the models encompass a wide
range of stellar effective temperatures from 55,000 K to 37,500 K, and
ionization parameters ($U=Q_{Ly}/(4\pi R^2~n_H~c)$) from 0.0006 to
0.06, the magnitude of the ICF is small overall and it shows a small
dispersion from model to model.  This suggests that it may be possible
to obtain reliable measurements of the C/O ratio from the C~II]
$\lambda$2326 and  [O~II] $\lambda$3727 lines even though C$^+$ and
O$^+$ are trace species in \HII\ regions, similar to [N~II]/[O~II] used
to derive N/O (Garnett 1990).

Stasi\'nska \& Leitherer (1996) provide a new set of photoionization
models using updated atomic data and expanding non-LTE stellar
atmospheres.  We show these results in the middle panel of Figure~6 for
stellar effective temperatures of 40,000, 50,000, and 60,000 K and a
range of ionization parameters.  Except for the coolest stars, the
models predict ionization correction factors which are in good
agreement with one another and with the Stasi\'nska (1990) models in
the top panel.  These results are again consistent with the suggestion
that the C$^+$/O$^+$ ratio can be used as a reliable C/O
indicator.

In the lower panel of Figure~6 we show the Stasi\'nska \& Leitherer
(1996) model results for an evolving burst of stars with an IMF of
slope $\alpha$=2.35.   Two metallicities, Z=0.1 Z$_\odot$ and Z=0.25
Z$_\odot$ are distinguished by line style.  Line style also
distinguishes models with different values of $FH$, the size of a
central evacuated hole in the spherical \HII\ region in units of the
Stromgren sphere radius.  Burst ages of 1, 2, 3, and 4 Myr are denoted
by different symbols along each evolutionary path. In all models, the
O$^{++}$ fraction drops sharply after 5 Myr when the most massive stars
have disappeared.  In models with Z=0.25 Z$_\odot$, the appearance of
Wolf-Rayet stars after $\sim$3 Myr produces an elevated He~II
$\lambda$4686 intensity (up to 0.05 that of H$\beta$) which is not
observed in these three objects.  For this reason, we do not include
points beyond 3 Myr for these models.  The starburst evolution shown in
the lower panel of Figure~6 is again consistent with the proposition
that X(C$^+$)/X(O$^+$) is well-behaved over a wide range in burst ages
and \HII\ region geometries.  For highly ionized nebulae,
0.75$<$X(O$^{++}$)$<$0.85, the reciprocal ionization correction factor,
X(C$^+$)/X(O$^+$), exhibits a small range from 0.6 to 0.8, although
this range is slightly offset from the single temperature models
discussed above.

Based on the photoionization models in Figure~6, we adopt an ionization
correction factor (ICF) of 1.2 in order to convert C$^+$/O$^+$ into the
C/O values summarized in Tables~4---6.  The small magnitude of this
correction may seem surprising since  C$^+$ is only a trace species in
\HII\ regions, X(C$^+$)$\approx$0.02---0.10.  At the same time, C$^+$
is the most abundant ionization state in photodissociation regions,
since the first ionization potential of C is only 11.3 eV.  However,
although C~II may be ubiquitous in the cooler environments exterior to
\HII\ regions, the excitation potentials of the UV $\lambda$2326 lines
are 5.3 eV, large enough that only in the hot, highly ionized
\HII\ regions will the electron temperature be high enough to
collisionally populate the upper levels.  Thus, as long as the C~II]
$\lambda$2326 and [O~II] $\lambda$3727 lines arise in the same physical
volume, their relative strengths provides a reasonablely accurate
estimation of the C/O ratio.

In summary, line blending with [O~III] $\lambda\lambda$2321,2331 and
uncertain ionization correction factors will ultimately limit the
usefulness of the C~II] $\lambda$2326 multiplet as an abundance
diagnostic.  However, in the absence of data on the higher ionization
species, it appears to be a useful substitute.  The small temperature
and reddening dependence of the O~III] $\lambda\lambda$1661,1666 to
C~III] $\lambda\lambda$1907,1909 ratio makes it the preferred method of
measuring C/O abundances (Garnett \etal\ 1995a).

\subsection{Abundance Results}

The chemical abundance ratios derived for each object appear in
Tables~4---6.  It is especially interesting to compare the C/O ratios
derived from C$^+$/O$^+$ versus those derived from C$^{++}$/O$^{++}$.
For NGC~4861 and T1345-420, 
we find that the C/O ratios estimated from C$^+$/O$^+$ are
nearly identical to those derived from
C$^{++}$/O$^{++}$.  For UM~469, the two estimates agree to within the
uncertainties due to reddening and electron temperature errors.
Furthermore, the C/O ratios abundances derived from C$^+$/O$^+$ appear
consistent {\it relative} to those derived from C$^{++}$/O$^{++}$ for
the three targets, supporting use of the C~II] $\lambda$2326
and [O~II] $\lambda$3727 lines as a valuable secondary 
diagnostics of the carbon abundance.

{\it NGC~4861}

The O$^{++}$ electron temperatures derived from measurements of [O~III]
$\lambda$4363 in the FOS and Calar Alto spectra are consistent,
14210$\pm$440 K and 14500$\pm$700 K respectively.  However, the ground
based longslit spectroscopy indicates substantial temperature
variations across the $\sim$20\arcsec region, ranging from 14,500 K
near the star cluster where the HST FOS aperture was centered, to
12,400 K in the nebular regions to the northeast.  The ground-based
spectra  reveal no compelling evidence for O or N/O abundance
variations across the $\sim$20\arcsec\ diameter nebular region at a
spatial resolution of 2\arcsec.  The derived O abundances are
12+log(O/H)=8.03$\pm$0.03 for the HST data, in excellent agreement with
12+log(O/H)=8.04$\pm$0.04 found by Garnett (1990).  The Calar Alto
spectra yield a slightly lower value, 12+log(O/H)=7.94$\pm$0.03, which
is consistent to within the errors.

The N abundance, log(N/O), as measured from the [N~II] $\lambda$6584
and [O~II] $\lambda$3727 lines is $-$1.40$\pm$0.07, which is typical
for dwarf galaxies of this metallicity, but slightly higher that the
value of $-$1.69$\pm$0.14 found by Garnett (1990).  The new value is
consistent with log(N/O)=$-$1.52$\pm$0.10 derived from the [N~II]
$\lambda$6584 and [O~II] $\lambda\lambda$7320,7330 lines in the Calar
Alto spectra.  Thus, we have confidence that the N/O ratios are
well-measured and are not much affected by reddening uncertainties.
The weighted mean N/O ratio used hereafter is log(N/O)=$-$1.44$\pm$0.06

The carbon abundance as derived from the C~III] $\lambda$1909 and
[O~III] $\lambda$5007 lines is log(C/O) = $-$0.51$\pm$0.10, consistent
with C/O ratios in other galaxies of similar metallicity (Garnett
\etal\ 1995a), and in excellent agreement with the C/O ratio derived
from the C~II] $\lambda$2326 and [O~II] $\lambda$3727 lines,
log(C/O)=$-$0.44$\pm$0.09.   However, given the uncertainties in
correcting for [O~III] contamination of the C~II] $\lambda$2326 line,
we adopt the C$^{++}$/O$^{++}$ result as the best-estimate value.
Since the electron temperature is well-measured, the largest source of
uncertainty on the N/O and C/O ratios is due to reddening.  The
uncertainty in c($H\beta$) dominates the quoted uncertainties in C/O
(and N/O where [O~II] $\lambda$3727 is used).  If a Galactic extinction
law instead of an SMC law is used to deredden the UV lines relative to
H$\beta$, the C/O ratio is decreased by $\sim$1.5$\sigma$, which still
places NGC 4861 among the other systems of similar metallicity.  Under
this extreme assumption, the resulting C abundance becomes log(C/O) =
$-$0.66$\pm$0.10 as derived from C$^{++}$/O$^{++}$ and log(C/O) =
$-$0.61$\pm$0.10 as derived from C$^{+}$/O$^{+}$.

From FOS measurements of the ultraviolet Si~III] 
$\lambda\lambda$1883,1892 and the C~III] $\lambda$1909 lines we measure
Si$^{++}$/C$^{++}$=0.126$\pm$0.040 using Equation 1 of Garnett
\etal\ (1995b).  The total elemental abundances, Si/C, can be estimated
using an ionization correction factor of $\sim$1.4 which is appropriate
for an O$^{++}$ fraction of $\sim$0.8 judging from photoionization
models summarized in their Figure~1.  The resulting total abundances
are Si/C=0.13$\pm$0.03 and log(Si/O)=$-$1.38$\pm$0.13.  This places
NGC~4861 among the objects with the highest Si/O ratio of those
measured by Garnett \etal\ (1995b).

{\it UM~469}

The O$^{++}$ electron temperatures derived from measurements of [O~III]
$\lambda$4363 are 12840$\pm$840 K and 12390$\pm$1000 K for the FOS and
Calar Alto spectra, consistent within the errors and in good agreement with
T$_e$=12590$\pm$1400 derived by CTM.  The derived O
abundances are marginally consistent with one another,
12+log(O/H)=7.97$\pm$0.05 for the HST FOS spectra versus 8.08$\pm$0.07
for the Calar Alto data.   Both compare favorably to
8.01$\pm$0.11 derived by CTM.  The N/O ratio, as
measured from the [N~II] $\lambda$6584 and [O~II] $\lambda$3727 lines,
is log(N/O)=$-$1.26$\pm$0.09 consistent with the CTM measurement,
log(N/O)=$-$1.19$\pm$0.10, placing it among the most N--rich objects at
this metallicity. These values compare favorably to
log(N/O)=$-$1.29$\pm$0.15 derived from the [N~II] $\lambda$6584 and
[O~II] $\lambda\lambda$7320,7330 lines, providing assurance that the N/O
ratio is genuinely high and is not adversely affected by reddening
uncertainties.

The derived C abundance is log(C/O)=$-$0.37$\pm0.22$, making it the
most C--rich object known at this metallicity.  The C/O ratio derived
from the C~II] $\lambda$2326 and [O~II] $\lambda$3727 lines is roughly
consistent at log(C/O)=$-$0.49$\pm$0.10.  If we were to adopt the
extreme assumption of a Galactic UV reddening law instead of the SMC
one, the resulting log(C/O) would be $-$0.65, placing UM~469 amidst all
of the other metal-poor galaxies at this metallicity (see Figures~8 \&
9).  The C/O ratio derived from the C~II] $\lambda$2326 and [O~II]
$\lambda$3727 lines dereddened with an Galactic type law is
log(C/O)=$-$0.61$\pm$0.10.  For UM~469, the uncertainties on the
reddening and the electron temperature contribute roughly equally to
the error budget for the C$^{++}$/O$^{++}$ measurement, while the low
S/N on the $\lambda$2326 line is the limiting factor for the
C$^{+}$/O$^{+}$ estimate.

From FOS upper limits on the ultraviolet Si~III] 
$\lambda\lambda$1883,1892 lines we measure Si$^{++}$/C$^{++}$$<$0.277.  Adopting an ICF
of 1.4 based on Figure~1 of Garnett \etal\ (1995b), the total elemental
abundances are Si/C$<$0.395 and log(Si/O)$<-$0.772.  These upper limits
are consistent with other metal-poor \HII\ regions, but are not very
helpful in constraining the Si abundance.

{\it T1345-420}

The O$^{++}$ electron temperature derived from measurements of [O~III]
$\lambda$4363 with the FOS is 12520$\pm$ 740 K.  This compares
favorably with 13320$\pm$ 720 K which we re-compute from the line
strengths given in CTM.  Likewise, the O abundances are consistent
within the errors, 12+log(O/H)=8.11$\pm$0.05 for the FOS data and
8.03$\pm$0.04 for the CTM data.  The log(N/O) derived from the CTM
spectra is $-$1.62$\pm$0.10, placing T1345-420 among objects with the
lowest known N abundances.

The C abundance derived from C~III] $\lambda$1909 and [O~III]
$\lambda$5007 lines is log(C/O)=$-$0.71$\pm$0.14, in excellent
agreement with log(C/O)=$-$0.69$\pm$0.12 from the C~II] $\lambda$2326
and [O~II] $\lambda$3727 lines.  The uncertainties on the reddening and
the electron temperature contribute roughly equally to the error budget
for the C$^{++}$/O$^{++}$ measurement, while the low S/N on the
$\lambda$2326 line is the limiting factor for the C$^{+}$/O$^{+}$
estimate.  In the extreme case that all of the intervening medium
follows a Galactic type reddening law (instead of the 50\% SMC, 50\%
Galactic proportion assumed above) the carbon abundances become
log(C/O)=$-$0.79$\pm$0.13 from C$^{++}$/O$^{++}$ lines versus
log(C/O)=$-$0.74$\pm$0.11 from the C$^{+}$/O$^{+}$ ratio.

From FOS upper limits on the ultraviolet Si~III] 
$\lambda\lambda$1883,1892 lines we measure Si$^{++}$/C$^{++}$$<$0.257.  Adopting an ICF
of 1.4 based on Figure~1 of Garnett \etal\ (1995b), the total elemental
abundances are Si/C$<$0.367 and log(Si/O)$<-$1.14.  These upper limits
are consistent with other metal-poor \HII\ regions, but are not very
helpful in constraining the Si abundances.

\section{C and N Abundances in Galaxies of Similar Metallicity}

These new data can begin to address which of the four scenarios
presented in the Introduction can best explain the variation in N/O
abundances among galaxies of a given metallicity.  Temporary N and He
enhancements from OB and Wolf-Rayet stars (i.e. ``pollution'', Pagel
\etal\ 1986; ``self-enrichment'', Kunth \& Sargent 1986)
 have been suggested as an explanation for the relatively high
abundances of these elements reported in some Wolf-Rayet galaxies
(Pagel \etal\ 1986, 1992; Esteban \& Peimbert 1995).   If N variations
among galaxies at constant O/H are all due to pollution from these
short-lived stars, then the abundance of carbon, produced predominantly
in long-lived, low to intermediate mass stars ($<$8 \mo; Timmes,
Woosley, \& Weaver 1995) should exhibit no correlation with the amount
of N enrichment.

In the case of NGC~5253, the two regions showing three-fold N
overabundances have identical C abundances, suggesting that N
production is decoupled from C production.  Since WR star winds are not
generally C-rich (except for the short-lived WC stars which become
increasingly rare in low-metallicity galaxies (see Arnault \etal\ 1989;
Maeder \& Maynet 1994), a scenario involving localized pollution from
the N-rich winds of 5--15 O and WR stars appears to be consistent with
both the magnitude of the N enrichment and the lack of observed C
enrichment (Walsh \& Roy 1989; Kobulnicky \etal\ 1997).  NGC~5253
appears to be an anomaly.  Except for the well-documented case in
NGC~5253, and possibly II~Zw~40 (Walsh \& Roy 1993) other nearby
starforming galaxies appear chemically homogeneous despite the presence
of multiple massive star clusters (NGC~1569: Devost, Roy, \& Drissen
1997; Kobulnicky \& Skillman 1997--- NGC~2366: Roy
\etal\ 1996---NGC~4214:  Kobulnicky \& Skillman 1996---NGC~4395: Roy
\etal\ 1996---NGC~6822:  Pagel, Edmunds, \& Smith 1980---the SMC and
LMC: Dufour \& Harlow 1977; Pagel \etal\ 1978; Russell \& Dopita
1990--- and assorted other irregular galaxies: Masegosa, Moles, \& del
Olmo 1991).  Galaxies with large populations of Wolf-Rayet stars do not
show systematically higher abundances than other actively star forming
systems (Kobulnicky \& Skillman 1996).

In the sample of galaxies presented here, there appears to be a
positive correlation between C and N abundances.  Such a correlation is
not expected if the N/O variations are due to nitrogen pollution from
short-lived massive stars.  In Figures 7 and 8 we plot the derived N/O and C/O
ratios versus metallicity, 12+log(O/H), for each of the three targets.
For comparison, we include the HST FOS results at 3 locations in NGC
5253, along with two other N/O and C/O measurements in objects with
similar O/H from the literature.  In Figure~7, T1345-420 occupies the
regime where the most N-poor objects are found.  NGC~4861 falls near
the middle of the distribution, while UM~469 lies among the most N-rich
objects, (excluding the extremely N-enriched regions of NGC~5253).  In
Figure~8, the three objects occupy the same relative positions along
the C/O axis, with UM~469 being the most C-rich and T1345-420 the least
C-rich.   The three new data points along
with previous measurements of C/O values in the SMC and NGC~2363 (Garnett
\etal\ 1995a) which have similar metallicity, help illustrate,
for the first time, the intrinsic dispersion in C/O.  However, the
uncertainties due to temperature and reddening errors are large, and
there are not yet enough data to characterize the magnitude of real
variations in C/O.

To examine the behavior of N and C abundances at a constant metallicity
(O/H), we plot in Figure~9 the N/O versus C/O ratios for all six
objects with available data in the narrow metallicity range,
8.0$<$12+log(O/H)$<$8.2.  Carbon abundance data for NGC~2363 and the
SMC N88A are taken from Garnett \etal\ (1995a) and the N abundances for
NGC~2363 are from Gonzalez-Delgado \etal\ (1994).  C/O ratios are derived
from C$^{++}$/O$^{++}$ in each case. The upper panel
shows the results of adopting the best-guess mixture of the Milky Way
and SMC type reddening laws for each object as described in \S\ 3.1.
The lower panel shows the resulting C/O ratios if we deredden UV lines
using only a Galactic-type law.  If NGC~5253 is ignored as a special
case of {\it localized} pollution, then there is a strong correlation
between the remaining 5 objects in the upper panel with a linear
correlation coefficient of 0.989.  This indicates that for 5 data
points selected randomly, the probability of getting an equal or
greater degree of correlation is 0.002.  A linear least-squares fit to
the data yields a slope of 2.47$\pm$0.84.   In the lower panel where a
Galactic reddening law is blindly assumed, the correlation coefficient
is 0.682, indicating that the probability of getting an equal or
greater degree of correlation is $\sim$0.20.  A linear fit to the data
yields a slope of 0.46$\pm$0.51. Given the small sample size and
substantial observational uncertainties, these data should not be used
to infer the form of a correlation between N/O and C/O.  We argue only
that, unless the extreme case of an entirely Galactic extinction
law is adopted, there is strong
evidence for a positive monotonic relationship between C and N
abundances among galaxies in this metallicity range.

In Figure~10 we plot the metallicity, as indicated by 12+log (O/H),
versus the C/N ratio for the three objects using the mean N/O and C/O
values listed in Tables 4---6.  The three galaxies observed in this
program show nearly identical C/N ratios, $<$log(C/N)$>$=0.933.  The
dispersion is 0.048, less than half the typical measurement error of
$\sigma_{C/N}$=0.15.  The newly measured value for NGC~4861 is
1.5$\sigma$ higher than that previously estimated using the IUE data
(Dufour \etal\ 1984).  The mean C/N value of log(C/N)=0.93 is in fair
agreement with, but on the high side of, the predictions of Matteucci's
(1986) model (a) for the solar neighborhood.  Excepting the N-enriched
locations in NGC~5253, most of the data in Figure~10 are consistent
with log(C/N)$\approx$0.90, though there is a tendency for lower C/N
ratios at lower metallicity.  Better C and N abundance measurements in
metal-poor systems are needed to establish whether the effective C/N
yields vary appreciably with metallicity.

\section{Interpretation of CNO Abundances}

The existence of a probable correlation in Figure~9, and the constant
C/N ratios at similar metallicity in Figure~10 implies that C and N
production mechanisms are coupled.  Such a correlation appears
inconsistent with the idea that massive stars are responsible for the
relative enrichment of either element.  Although C may be produced in
massive stars and released in supernova ejecta, the majority of C in
galaxies is thought to originate in low to intermediate mass stars 
($<$8 \mo; Timmes \etal\ 1995).  Furthermore, although some N may be produced in the
atmospheres of massive stars and released in massive star winds, little
or no N is released in supernovae (see stellar evolution and
nucleosynthesis models of Schaller \etal\ 1992; Woosley \& Weaver
1995).  The correlation in Figure~9 is consistent with the hypothesis
that both the N and C abundances are the result, not of local, temporary
enrichment from massive stars, but of global, secular enrichment
resulting from the particular star formation history of low and
intermediate mass stars in each object.

The current data are not able to address whether the N is predominantly
primary or secondary origin, since intermediate mass stars may produce
either or both types of N (Renzini \& Voli 1981).  Nor are we able to
distinguish between the latter three hypotheses outlined in the
introduction.  That is, without additional information on the star
formation histories and burst ages in each object, we are unable to say
whether the observed N/O and C/O variations are due to 1) variations in
the IMF from one galaxy to the next, 2) a time delay between the release
of N \& C and the release of O, or 3) preferential loss of O due to
galactic winds in galaxies with high N/O and C/O ratios,
or some combination of the three options.  However, it
is possible to make some predictions within the context of each
hypothesis.  Observational work verifying one or more of these
predictions would help to distinguish between the three hypotheses.

{\it The Variable IMF Hypothesis}

The correlation between C and N among these galaxies is revealing
because it seems to confirm that the N/O and C/O variations among our
sample are due to variations in the effective N and C yields from low
and intermediate mass stars compared to the O yields from massive
stars.  N is produced predominantly from C and O in the
CNO cycle, and may be either ``primary'' or ``secondary'' in origin.
Primary nitrogen is synthesized directly from H and He via fresh C and
O while secondary nitrogen is made from pre-existing C and O.  Primary
N is thought to be produced in low mass (4-5 \mo) stars (Renzini \&
Voli 1981), or possibly in massive stars $>$30 \mo\ (Woosley \& Weaver
1995).  Secondary N is thought to be produced in intermediate mass
stars and released into the interstellar medium (ISM) through red giant
winds and planetary nebulae.  The ratio of primary to secondary N
produced in massive versus intermediate mass stars is presently poorly
constrained both observationally and theoretically.   Consequently, the
expected relation between the abundance of N and other elements within
the ISM is poorly known.  If nitrogen is predominantly of primary
origin, the nitrogen abundance is expected to show a linear correlation
with the oxygen abundance, while if N is mostly of secondary origin, N
should be proportional to the square of the oxygen abundance.  However,
the exact ratio of N to other elements depends upon many factors,
including the fraction of N from primary and secondary nucleosynthesis,
the initial metallicity of the stars, and perhaps most importantly, on
the stellar initial mass function.

In the context of the variable IMF hypothesis, we predict that
UM~469 should have a steeper IMF (greater fraction of low mass stars)
than the other systems in order to produce substantially more N and C
relative to O.   Since it is very difficult to measure the initial mass
function even in nearby systems, obtaining suitable measurements for
the very distant UM~469 is probably not feasible.  However, there is no
compelling evidence that the initial mass function varies in local
galaxies (Hunter \etal\ 1997; Parker \& Garmany 1993; Hill, Madore, \&
Freedman 1994; but for an opposing view see Charlot \etal\ 1993), so
this hypothesis is not especially attractive at the present time in the
absence of new results providing strong evidence for IMF variations.

{\it The Delayed Release or ``Clock'' Hypothesis}

Another way to obtain varying C/O and N/O ratios at constant
metallicity while maintaining a universal IMF and universal stellar
yields is to delay the release of N and C relative to O (Edmunds \&
Pagel 1978; Garnett 1990).  Models of chemical evolution in galaxies which produce
stars in bursts separated by long quiescent periods (Edmunds \& Pagel
1978; Pantelaki 1988; Clayton \& Pantelaki 1993) suggest that the
dispersion in N/O could be due to a delayed relase of N and elements
produced in low--mass longer lived stars, compared to O and elements
produced in massive, short-lived stars.

The Delayed Release Hypothesis  predicts that the N/O
ratio evolves significantly during a single cycle of star formation,
followed by quiescence.  Figure~11 illustrates this cycle
schematically.  During a long period of quiescence, the evolution of
intermediate mass stars should enrich a galaxy significantly in N (and,
we suggest, carbon too), but not O or any of the products of Type~II
supernovae.  At the beginning of a burst of SF which occurs after a
long period of quiescence, the N/O ratio should be high as the result
of the evolution of intermediate mass stars over the last few 100 Myr.
After the quiescent period ends, and massive star formation commences,
the N/O ratio drops and the O/H ratio increases as supernovae release O
and other $\alpha$ process elements into the ISM.  The chemical
properties of galaxies should then evolve rapidly during the few tens
of millions of years after the start of a massive starburst that
dominates the host galaxy.    At the end of a period of massive star
formation, the N/O ratio should be a minimum as the massive stars die
and the \HII\ region fades away.

If this picture is correct, the N/O ratios could be used as a ``clock''
to determine the time since the last major episode of star formation.
Since He is often  produced along with N, the He/H ratios could be
affected in the same way, although variations will be more difficult to
detect against a pre-existing background of primordial He.  Based on
data presented here, the C/O ratios may vary in roughly the same way
as N/O during the course of a starburst.  Curiously, II~Zw 40 and NGC 5253
appear to fit well into the predictions of the Delayed Release
Hypothesis, showing high N/O ratios and evidence for extremely young
starbursts  based on large EW(H$\beta$) and completely thermal radio
continua (see Deeg \etal\ 1993; Beck \etal\ 1996; Rieke, Lebofsky, \&
Walker 1988).  The chemical properties of the Pegasus dwarf irregular
galaxy also seem consistent with this scenario (Skillman, Bomans, \&
Kobulnicky 1997).

Generally, the Delayed Release Hypothesis  predicts that
\HII\ and blue compact galaxies with high N/O ratios (UM469, for
example) are experiencing their first burst of massive star formation
after a relatively long quiescent interval, while the same types of
galaxies with low N/O ratios (T1345-420, NGC~6822) have had little or
no quiescent interval.  However, since the nucleosynthetic products of
massive stars appear to require longer than $\sim$10$^7$ years to mix
with the surrounding ISM and become detectable (Tenorio-Tagle 1996;
Kobulnicky \& Skillman 1997 and references therein), there will be some
time lag ($>$10$^7$ yrs) between the supernova explosions and the
appearance of fresh $\alpha$-process elements in the warm ionized ISM.
As long as this time lag is not longer than the lifetimes of N and
C-producing stars ($\sim$10$^8$ yrs for 5 \mo\ stars; we assume that C
and N released in cool, slowly-expanding red giant winds and planetary
nebulae do not require such long timescale to mix with the surrounding
medium), then the N/O ratio may serve as a useful ``clock'' of the time
elapsed since the last major SF episode.  If, however, the time lag
required for massive stars ejecta to cool and mix with the ISM is
longer than $\sim$10$^8$ yrs, then the N/O ratio is unlikely to
accurately reflect the time since the most recent starburst.

Further observational or theoretical work showing that the timescales
for cooling and mixing of hot supernova ejecta from massive
star clusters are $<$10$^8$ (but $>$10$^7$ years implied by the lack of
observed self-enrichment in \HII\ regions) years would add plausibility
to the clock hypothesis.  A further test of the clock hypothesis will
come from measuring star formation histories in galaxies spanning a
range of N/O ratios.  Unfortunately, all of the objects in this sample
except NGC~4861 are too far away to directly investigate their star
formation histories from stellar color-magnitude diagrams.  Infrared
imaging of the underlying older stellar population could, in principle,
provide an estimate of the relative ages of each burst.

{\it The Differential Mass Loss Hypothesis}

Under this hypothesis which involves differential heavy element loss in
galactic winds, the IMF is universal, constant in time and the N/O
ratio for a given galaxy does not change with time as long as the {\it
fraction} of heavy elements that are lost in galactic winds remains
constant.  The hypothetical role of metal-enriched galactic winds has
been discussed extensively (Mathews \& Baker 1971; Dekel \& Silk 1986;
Vader 1987; De Young \& Gallagher 1990).  Preferential loss of oxygen
and Type II SNe products seems required in some cases to avoid
``overproducing'' oxygen in low mass galaxies (Marconi, Matteucci, \&
Tosi 1994; Esteban \& Peimbert 1995) but observational evidence
(Marlowe \etal\ 1995 and references therein) for their impact on the
evolution of galaxies is still inconclusive.

In the context of this fourth hypothesis, we predict that UM~469, which
has the highest N/O and C/O ratios, has suffered the strongest O loss,
while T1345-420, with the lowest N/O and C/O ratios, retains metals
produced in supernovae most efficiently of the three.  Velocity
resolved spectroscopy of the ionized gas in each object could address
whether galactic winds or ``blowout'' are {\it presently} occurring.
H$\alpha$ or \HI\ mapping would provide estimates of important
parameters which affect the mass loss rates in galaxies, such as the
gas content and potential well depth.  Given the difficulty in
establishing the magnitude of galactic winds even in local galaxies, it
is unlikely that the distant UM~469 could be investigated in this
manner.  The nearer objects, however, are prime targets for further
observations.

\section{Conclusions}

New measurements of the C abundance in a sample of three starforming
galaxies with 12+log(O/H)$\approx$8.1 reveal evidence of a correlation
between the N/O and C/O ratios.  Objects with higher N/O also have
higher C/O, consistent with the idea that C and N production
nucleosynthesis is dominated by low to intermediate mass stars.  The
slope of the relationship is presumably fixed by the IMF which dictates
the ratio of low-mass C-producing stars to intermediate mass stars
which are thought to dominate N production.  The observed correlation
is inconsistent with the chemical ``pollution'' scenario whereby
massive stars and Wolf-Rayet winds create localized N enhancements,
{\it unless} massive stars also produce proportionally large localized
C enhancements, contrary to theoretical expectations.  Although we
reject localized chemical pollution from massive stars as an
explanation for the scatter in N/O and C/O at a given metallicity, we
outline three additional hypotheses capable of explaining the data, and
we suggest approaches to discriminate between them.  At the present
time, these conclusions are based upon only 5 objects, and should be
treated as preliminary.  More C abundance measurements in galaxies with
high N/O ratios and 12+log(O/H)$\approx$8.1 are needed to establish
strength and form of the C/O versus N/O relationship.

We have also presented a comparison of the C/O ratios in \HII\ regions
derived independently from the C$^{++}$/O$^{++}$ ratios (via C~III]
$\lambda$1909 and [O~III] $\lambda$5007 lines) and from the
C$^{+}$/O$^{+}$ ratios (via C~II] $\lambda$2326 and [O~II]
$\lambda$3727 lines).  Both approaches yield C/O ratios that agree to
within the formal uncertainties. Considering that C$^+$ is a trace
species in \HII\ regions, and that C~II] $\lambda$2326 is blended with
the [O~III] $\lambda\lambda$2321,2331 lines at low spatial resolutions,
the good agreement between C$^{++}$/O$^{++}$ and C$^{+}$/O$^{+}$
diagnostics suggests that this method may be useful for determining
carbon abundances in emission-line galaxies when data on the dominant
ionization species, C$^{++}$ and O$^{++}$, are not available.     Both
approaches are limited primarily by uncertainties on the measured
electron temperature and uncertainties in the extinction law used to
deredden the ultraviolet lines.  More precise measurements of the C/O
ratios in galaxies can be made using the O~III] $\lambda$1666/C~III]
$\lambda$1909 line ratio which is less sensitive to reddening and
temperature errors than the method used here, but this generally
requires objects with high nebular surface brightness and/or an
instrument with excellent UV sensitivity.  Although C~II] $\lambda$2326
is typically quite weak in extragalactic \HII\ regions, observations of
this multiplet, along with [O~II] $\lambda$3727, can, in principle, be
used to assess the C/O ratios of high-redshift emission line objects
out to $z\approx1.5$.

\acknowledgments  We would like to thank Robbie Dohm-Palmer and Sabina
M\"ohler and the Calar Alto staff for assistance at the Calar Alto 3.5
m telescope.   We are grateful for comments and suggestions from Don Garnett,
Bernard Pagel, Greg Shields, and a very thorough review
of the draft
manuscript by Reggie Dufour.  
H.~A.~K. is grateful for financial assistance from a NASA
Graduate Student Researchers Program fellowship.  H.~A.~K.  and
E.~D.~S. were supported by NASA LTSARP Grant No.  NAGW--3189.   Further
support for this work was provided by NASA through grant number
GO-6801.01-95A from the Space Telescope Science Institute, which is
operated by the Association of Universities for Research in Astronomy,
Inc, under NASA contract NAS5-26555.

\clearpage

\begin{figure}
\plotone{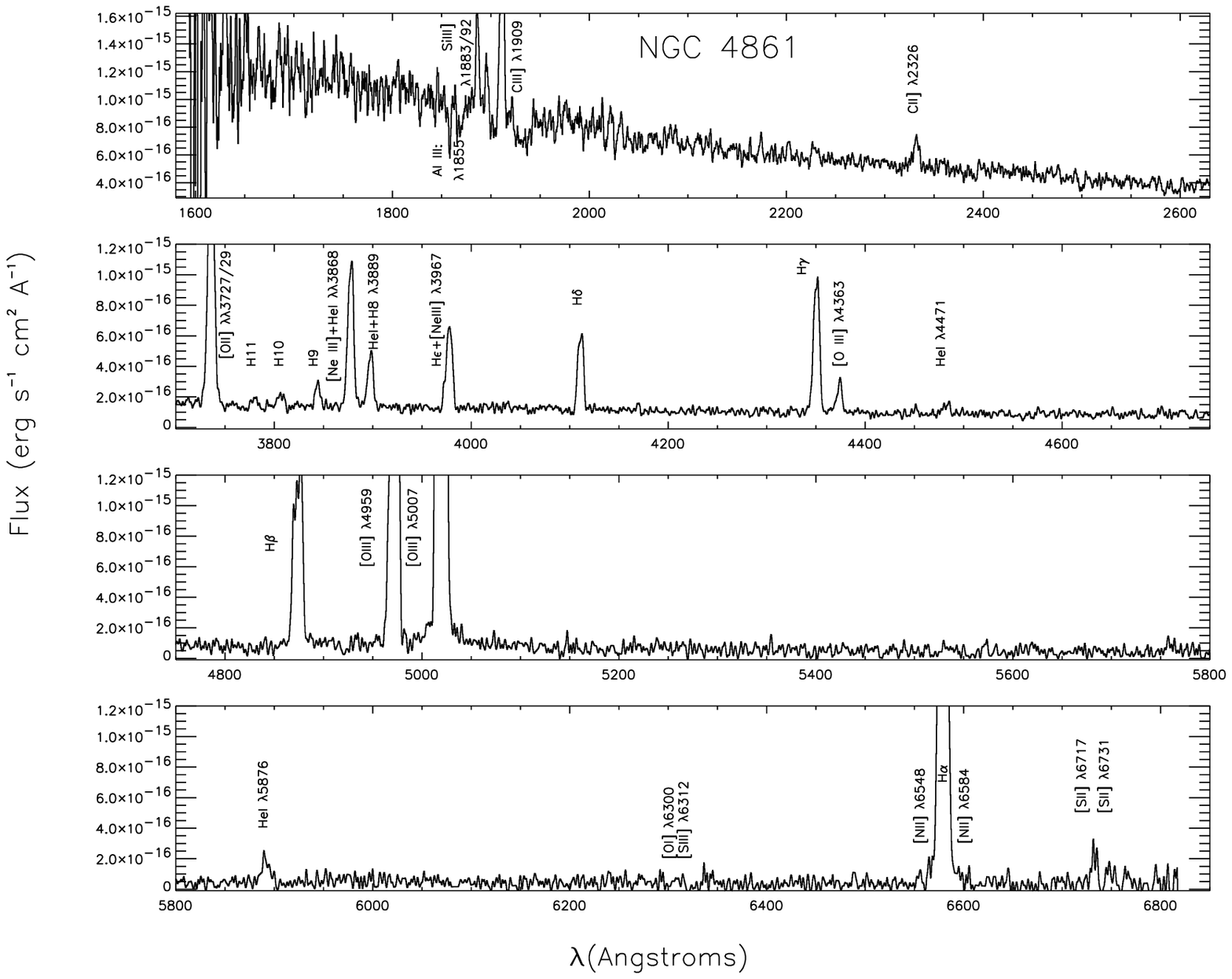}
\figcaption[.ps] {HST FOS spectrum of NGC~4861.  Four gratings allow a
wavelength coverage from 1200 \AA\ to 6800 \AA.  A strong
[O~III] $\lambda$4363 line provides an accurate electron temperature
determination while detections of and C~III] $\lambda$1909 and and
C~II] $\lambda$2326 allow measurements of the $C^{++}/O^{++}$ and
$C^{+}/O^{+}$ ratios.  Table~2 lists the dereddened line strengths.
\label{FOS-N4861} }
\end{figure}

\begin{figure}
\plotone{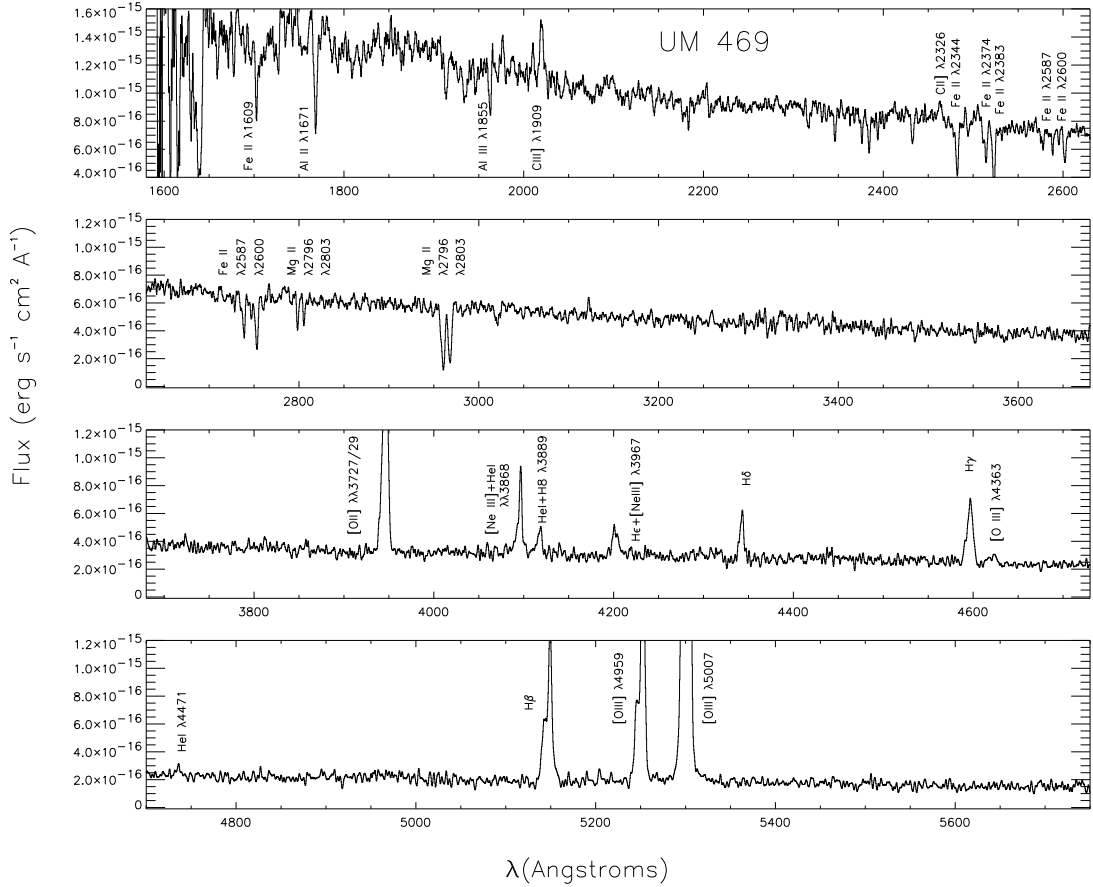}
\figcaption[.ps]
{HST FOS spectrum of NGC~UM469.  Four gratings allow
a wavelength coverage from 1200 \AA\ to 6800 \AA, similar to Figure~1.  
H$\alpha$ and the [N~II] $\lambda\lambda$6548, 6584 lines
are redshifted out of the G570H passband, so reddenings
are estimated from the higher order Balmer line ratios.
Table~2 lists the dereddened line strengths.
Some prominent interstellar absorption lines that arise within
the Galaxy and within UM~469 are labeled at their respective redshifts.
\label{FOS-UM469} }
\end{figure}

\begin{figure}
\plotone{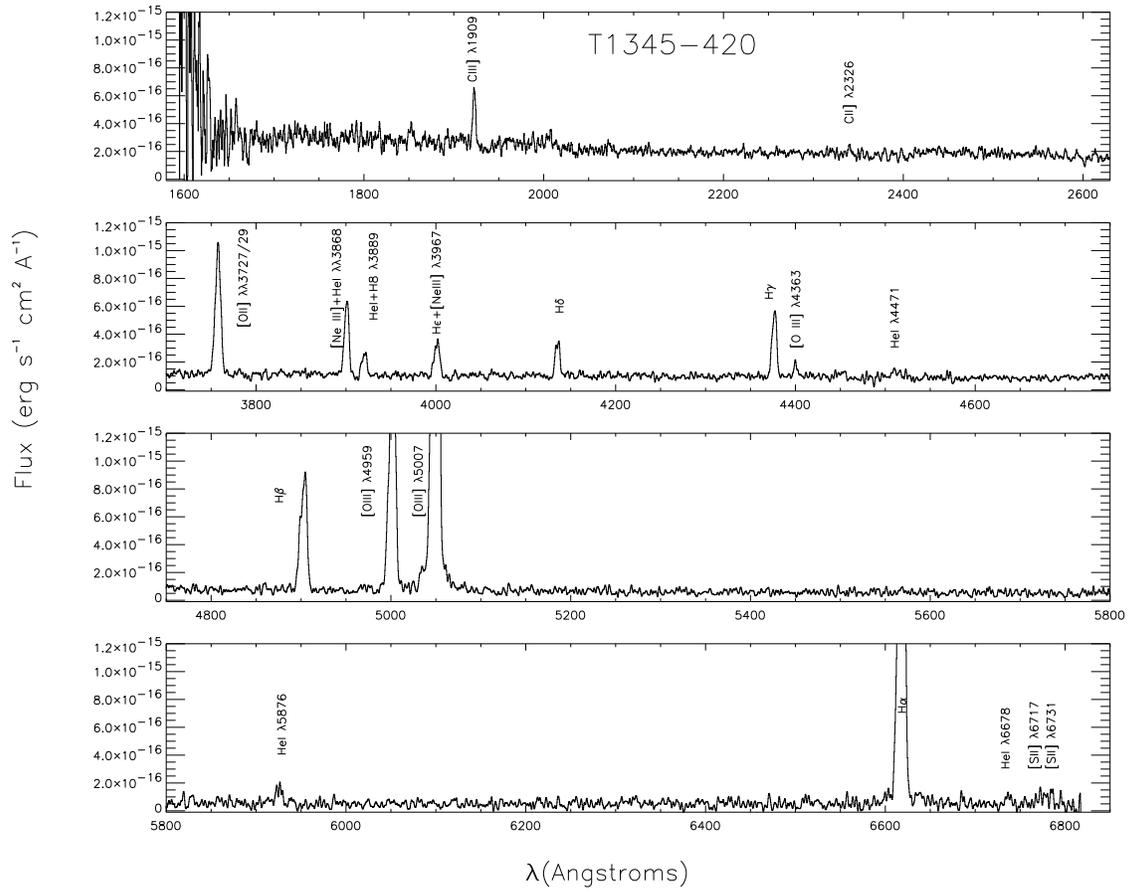}
\figcaption[.ps] {HST FOS spectrum of T1345-420.  Wavelength coverage
ranges from 1200 \AA\ to 6800 \AA, similar to Figure~1.  Table~2 lists
the dereddened line strengths.  \label{FOS-T13451} }
\end{figure}

\begin{figure}
\plotone{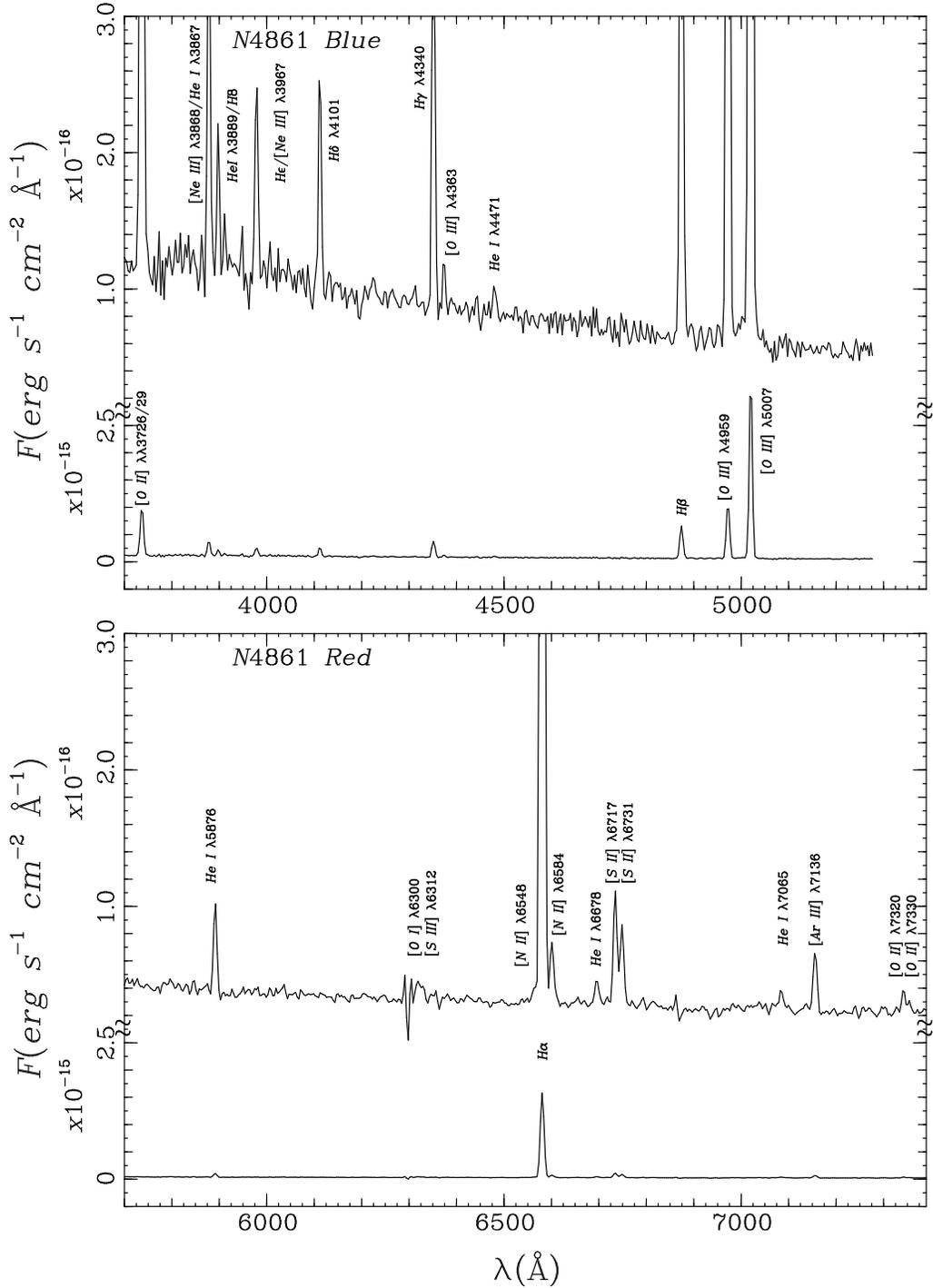}
\figcaption[.ps] {Calar Alto 3.5 m twin spectrograph spectrum of
NGC~4861 from which we measure the N/O ratio.  Wavelength coverage
ranges from 3560--5280 \AA\ in the blue and 5640--9170 \AA\ in the
red.  Detections of the [O~III] $\lambda$4363 allow an  accurate
measure of the electron temperature.  Table~3 lists the dereddened line
strengths.  \label{CA-N4861} }
\end{figure}

\begin{figure}
\plotone{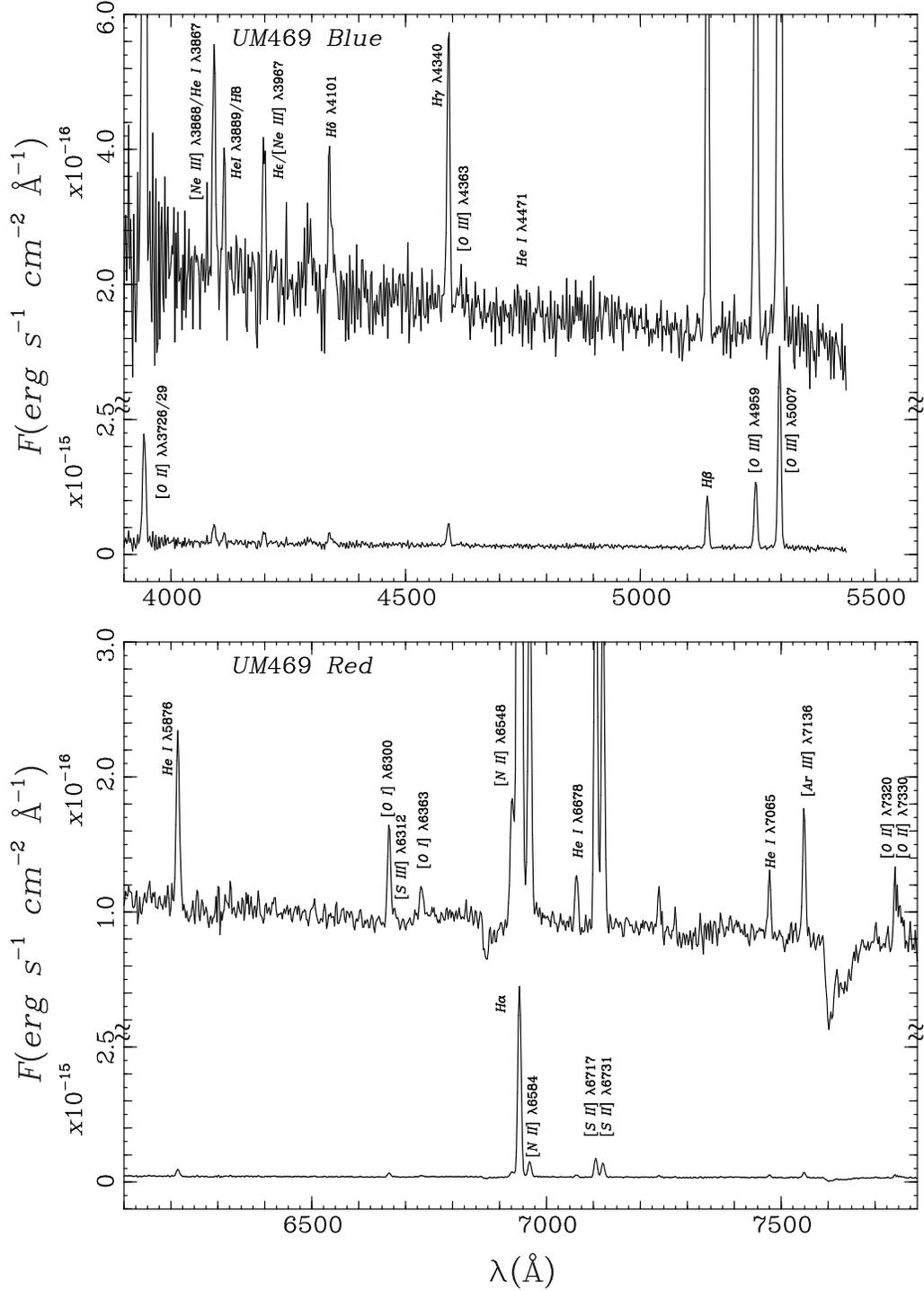}
\figcaption[.ps]
{Calar Alto 3.5m twin spectrograph  spectrum of UM~469.   
Wavelength coverage ranges from 3400--5400 \AA\ in
the blue and 5700--9600 \AA\ in the red.
A direct measurement of the electron temperature 
using the 4$\sigma$ detection of [O~III] $\lambda$4363 
is consistent with the HST FOS result.
Table~3 lists the dereddened line strengths.
\label{CA-UM469} }
\end{figure}

\begin{figure}
\centerline{\psfig{file=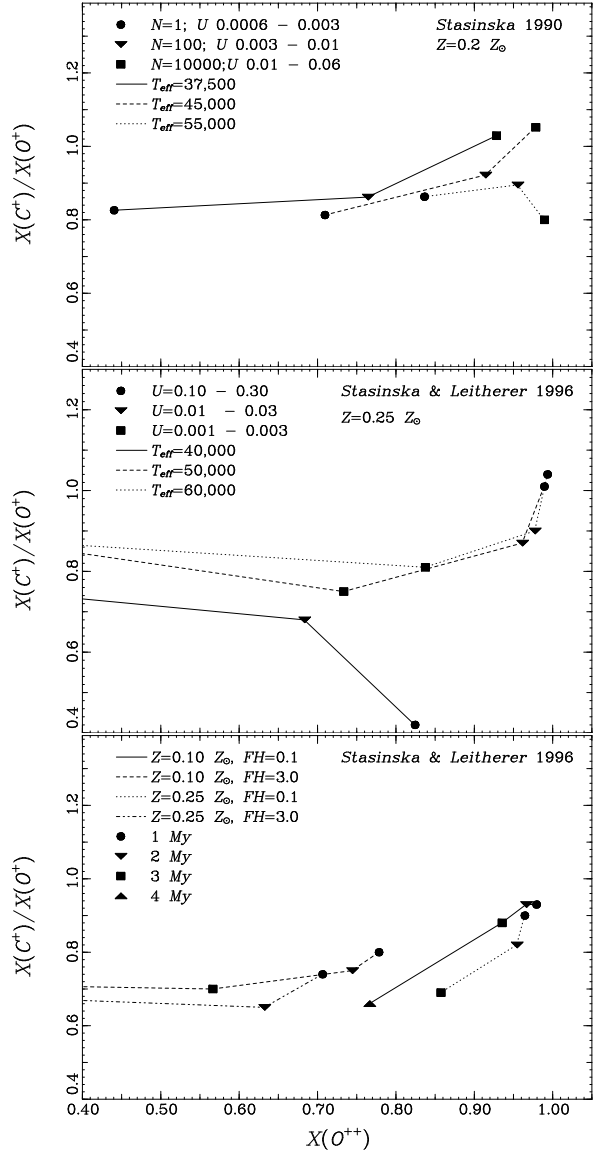,width=3.2in}}
\figcaption[.ps] {Photoionization models of X(C$^+$)/X(O$^+$) versus
the fraction of doubly ionized oxygen, X(O$^{++}$).
[X(C$^+$)/X(O$^+$)]$^{-1}$ is the ionization correction factor that
converts the observed C$^+$/O$^+$ ratio into C/O.  A wide range of
model parameters is explored.  Top panel:  Lines connect models
(Stasi\'nska 1990) with the same temperature stars, while symbols
denote models with similar ionization parameter.  Middle panel:
\HII\ region models (Stasi\'nska \& Leitherer 1996) with the stars of
identical temperature.  Symbols denote models with similar ionization
parameters while lines connect models of similar stellar temperature.
Lower panel: Photoionization models for an evolving cluster of stars
with different ages and metallicities.  See text.  The models
demonstrate that for a wide range of \HII\ region parameters the
fraction X(C$^+$)/X(O$^+$) is consistent and has a small dispersion
(0.65 to 0.9) when the O$^{++}$ fraction is in the range
0.5$<$X(O$^{++}$)$<$0.85.  This suggests that measurements of the C~II]
$\lambda$2326 multiplet and [O~II] $\lambda\lambda$3727,3729 lines can
be used as a reliable indicator of the C/O ratio.  \label{diag} }
\end{figure}

\begin{figure}
\centerline{\psfig{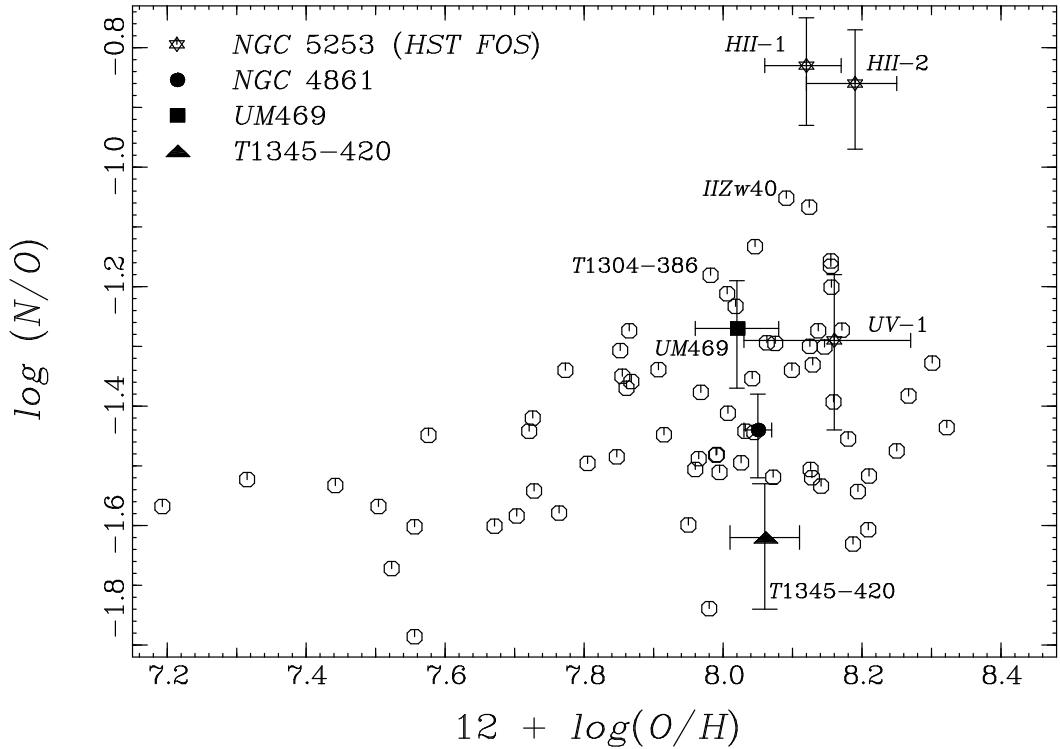}}
\figcaption[.ps] {12+log(O/H) versus log(N/O) for NGC 4861, UM~469, and
T1345-420.  Included for comparison are 59 other metal-poor systems
from the compilation of Kobulnicky \& Skillman (1996) and 3 locations
within NGC~5253 (Kobulnicky \etal\ 1997).  UM~469 is among the most
N--rich objects for its metallicity, excepting NGC~5253 and II~Zw~40
which exhibit strong N enhancements.  NGC~4861 has a fairly typical N/O
ratio for its metallicity, while T1345-420 is among the lowest.
\label{OH_NO} }
\end{figure}

\begin{figure}
\centerline{\psfig{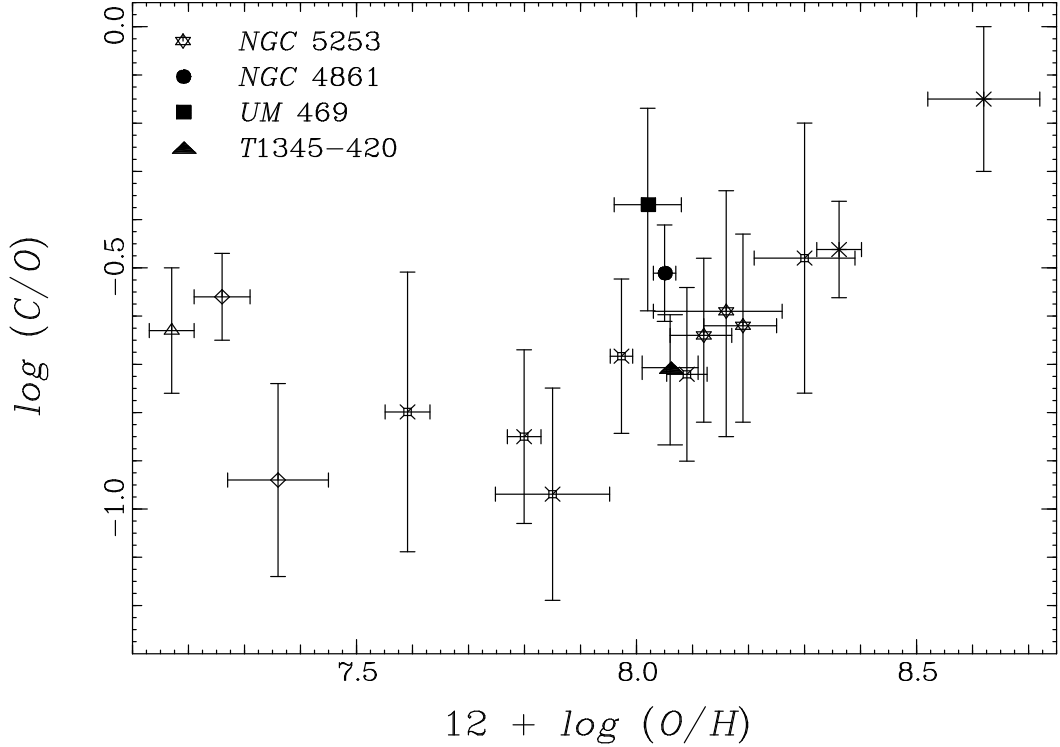}}
\figcaption[.ps] {12+log(O/H) versus log(C/O) for NGC 4861, UM~469, and
T1345-420.  Filled symbols mark C/O ratios derived from
C$^{++}$/O$^{++}$ while open symbols of the same type mark C/O ratios
computed from C$^{+}$/O$^{+}$.  Included for comparison are 11 other
metal-poor systems from Garnett \etal\ 1997 (I~Zw~18; diamonds),
Garnett \etal\ 1995a (crosses with circles), and Dufour 1984
(crosses).  Three independent locations within NGC~5253 (Kobulnicky
\etal\ 1997) are also plotted as stars.  The C abundance in UM~469
exceeds that of systems with similar metallicity, including the two
N--overabundant positions within NGC~5253, although the uncertainties
due to reddening are significant.  NGC~4861 and T1345-420 have fairly
typical C/O ratios for their metallicity.  \label{OH_CO} }
\end{figure}

\begin{figure}
\plotone{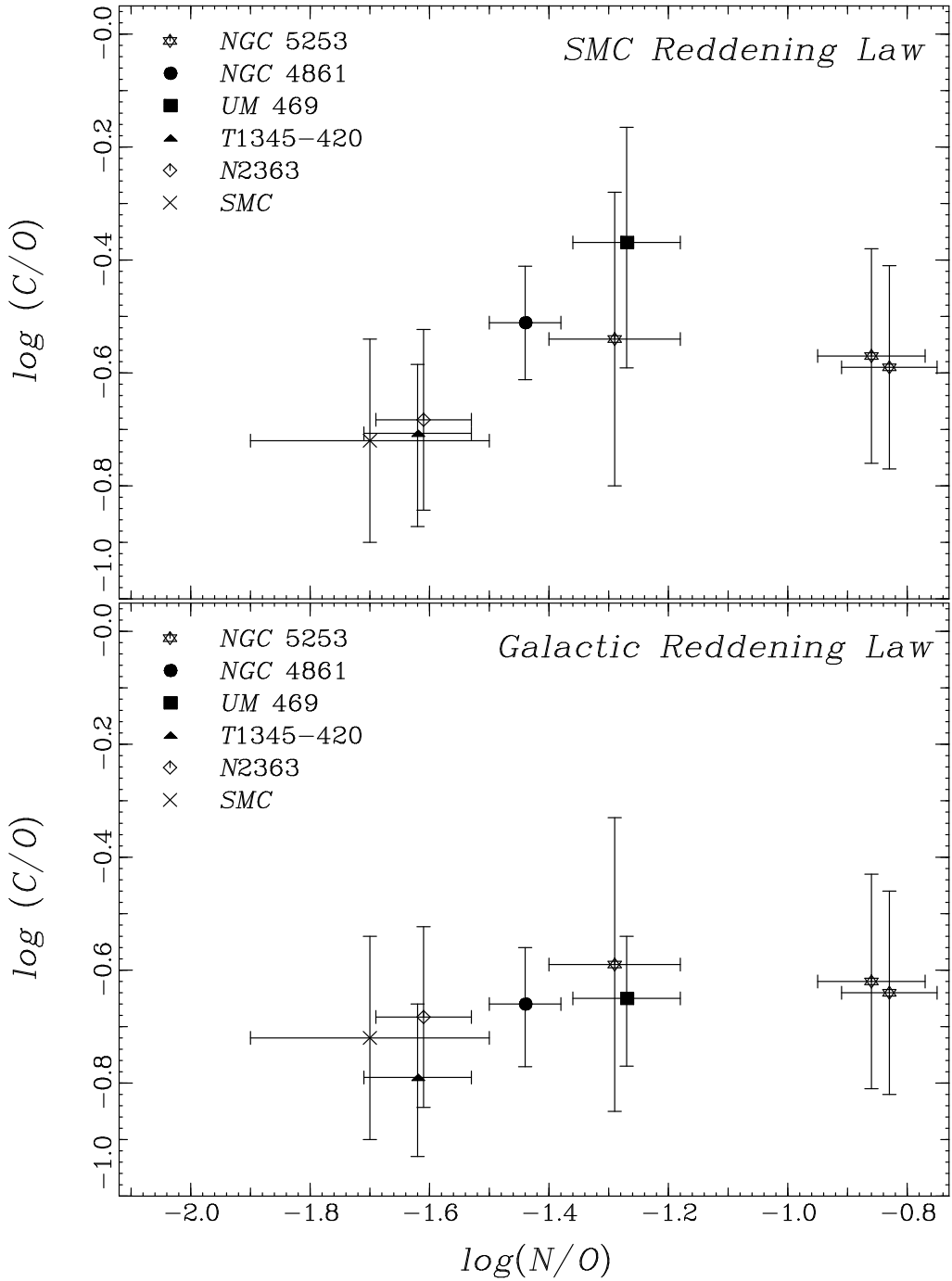}
\voffset -0.3in
\figcaption[.ps] {log(N/O) versus log(C/O) for galaxies with measured N
and C abundances and metallicity in the range 8.0$<$12+log(O/H)$<$8.2.
Top panel: The C/O ratios plotted are derived from  C$^{++}$/O$^{++}$
 as described in the text assuming that extinction within the
metal-poor systems follows an SMC type reddening law.  Lower panel: The
C/O ratios are derived from C$^{++}$/O$^{++}$
 as described in the text but assuming that extinction within the
metal-poor systems follows a Galactic type reddening law.  Excepting
the two nitrogen-polluted locations within NGC~5253, the data are
consistent with a positive monotonic relationship between the C and N
abundances.  However, the uncertainties are large, and the form of the
relation is poorly constrained.  \label{NO_CO} }
\end{figure}

\begin{figure}
\centerline{\psfig{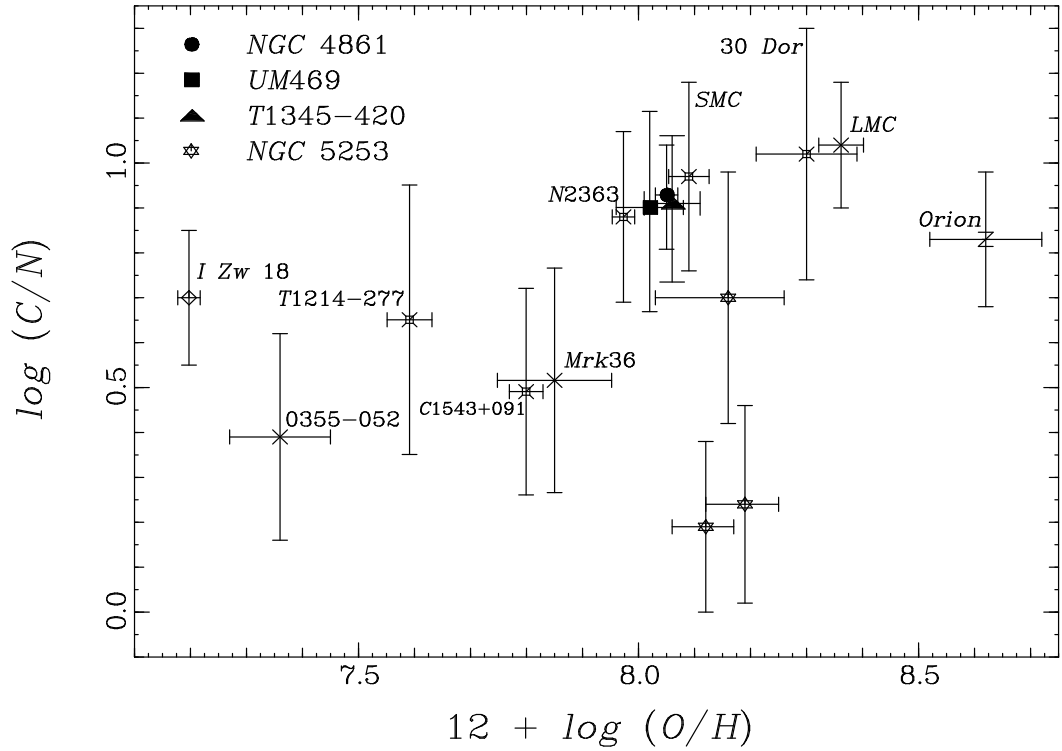}}
\figcaption[.ps] {12+log(O/H) versus log(C/N) for galaxies with
measured N and C abundances and 8.0$<$12+log(O/H)$<$8.2.  The mean
value of log(C/N)=0.96 for NGC~4861, UM~469, and T1345-420 is in fair
agreement with the C/N predictions of Matteucci's (1986) model (a) for
the solar neighborhood.  Based on present data, there is weak evidence
for lower C/N ratios among the most metal-poor objects.
 \label{OH_CN} }
\end{figure}

\begin{figure}
\centerline{\psfig{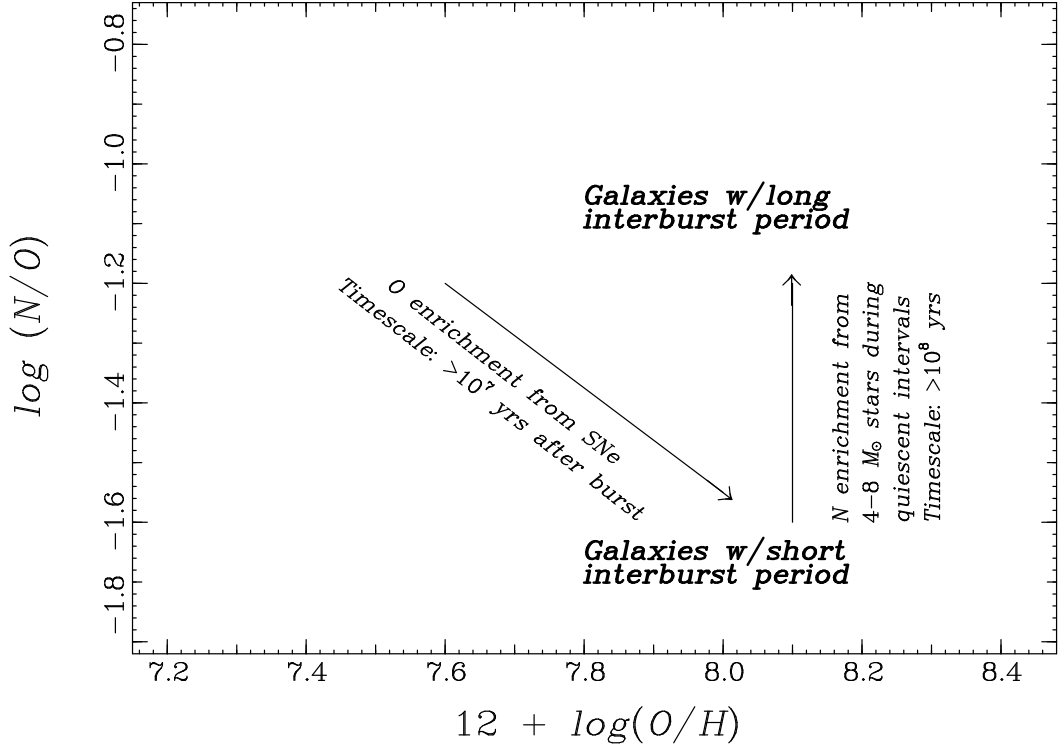}}
\figcaption[.ps] {Schematic O/H vs. N/O diagram showing how the N/O
ratio is expected to evolve during the course of a complete
starburst-quiescent cycle.  In the context of the ``clock''
hypothesis,  galaxies with large N/O ratios are experiencing their
first starburst after a long quiescent interval, whereas galaxies with
low N/O ratios had a short or no quiescent period since the previous
major star formation episode.   The observed correlation between the
N/O and C/O ratios suggests that the C/O ratio might be a useful
indicator of the interval since themost recent star formation episode
as well.
 (see also Edmunds \& Pagel 1978; Pantelaki 1988; Garnett 1990;
Pilyugin 1992; Clayton \& Pantelaki 1993.)
 \label{schem} }
\end{figure}

\end{document}

\begin{planotable}{lrrr}
\tablecaption{HST FOS Observation Log}
\tablewidth{270pt}
\tablehead{
\colhead{Object} & 
\colhead{grating}   & 
\colhead{resolution}   & 
\colhead{exposure}   \\ [.2ex]  
\colhead{Date}   & 
\colhead{}   & 
\colhead{(\AA) }   & 
\colhead{(sec)}    
 }
\startdata
NGC 4861 &   G190H   & 3 & 2370 \nl
1997 Jan 28 &   G270H   & 4.5 & 1200 \nl
         &   G400H   & 6 &  400 \nl
         &   G570H   & 9 &  180 \nl
UM469    &   G190H   & 3 & 3840 \nl
1996 Jul 8  &   G270H   & 4.5 & 1800 \nl
         &   G400H   & 6 &  960 \nl
         &   G570H   & 9 &  700 \nl
T1345-420&   G190H   & 3 & 2440 \nl
1996 Aug 16  &   G270H   & 4.5 & 1400 \nl
         &   G400H   & 6 & 790 \nl
         &   G570H   & 9 & 410 \nl
\enddata
\end{planotable}
\clearpage

\begin{planotable}{lrrrrr}
%\tablecaption{HST FOS Dereddened Line Strengths  \label{lines} }
\tablecaption{HST FOS Dereddened Line Strengths}
\tablewidth{480pt}
\tablehead{
\colhead{Line} & 
\colhead{f($\lambda$)}   & 
\colhead{f($\lambda$)}   & 
\colhead{NGC 4861\tablenotemark{a}}   & 
\colhead{UM469\tablenotemark{a}}   & 
\colhead{T1345-420\tablenotemark{a}}   \\ [.2ex]  
\colhead{  }   & 
\colhead{MW}   & 
\colhead{SMC}   & 
\colhead{${I_0}\over{I_{H\beta}}$ }   &
\colhead{${I_0}\over{I_{H\beta}}$ }   &
\colhead{${I_0}\over{I_{H\beta}}$ }    }
\startdata
1661 O \III]  & 1.08 & 2.18 & $<$0.631\tablenotemark{b} & $<$1.027\tablenotemark{b} & $<$0.457\tablenotemark{b}\nl  
1750 N \III]  & 1.00 & 1.94 & $<$0.565\tablenotemark{b} & $<$0.842\tablenotemark{b} & $<$0.422\tablenotemark{b} \nl  
1883 Si \III] & 0.91 & 1.71 & 0.384$\pm$0.038  & $<$0.550\tablenotemark{b}          & $<$0.265\tablenotemark{b} \nl  
1892 Si \III] & 0.91 & 1.71 & 0.254$\pm$0.025  & $<$0.550\tablenotemark{b}          & $<$0.265\tablenotemark{b} \nl  
1909 C \III]  & 0.90 & 1.66 & 1.401$\pm$0.360  & 0.833$\pm$0.401                    & 0.466$\pm$0.101 \nl  
2321 [O \III]\tablenotemark{c} & 0.75 & 1.08 & 0.028$\pm$0.004  & 0.020$\pm$0.005   & 0.009$\pm$0.002 \nl  
2326 [C \II]\tablenotemark{c}  & 0.75 & 1.08 & 0.140$\pm$0.022  & 0.120$\pm$0.031   & 0.045$\pm$0.011 \nl  
3727 [O \II]  & 0.25 & 0.25 & 1.195$\pm$0.112  & 1.344$\pm$0.162                    & 1.043$\pm$0.100 \nl  
3869 [Ne \III]& 0.22 & 0.22 & 0.612$\pm$0.052  & 0.318$\pm$0.037                    & 0.502$\pm$0.045 \nl  
4101 H$\delta$& 0.18 & 0.18 & 0.315$\pm$0.022  & 0.273$\pm$0.027                    & 0.257$\pm$0.021 \nl  
4343 H$\gamma$& 0.12 & 0.12 & 0.565$\pm$0.027  & 0.468$\pm$0.029                    & 0.450$\pm$0.024 \nl  
4363 [O \III] & 0.11 & 0.11 & 0.119$\pm$0.008  & 0.055$\pm$0.009                    & 0.076$\pm$0.012 \nl  
4471 He \I    & 0.09 & 0.09 & 0.040$\pm$0.006  & 0.056$\pm$0.009                    & 0.048$\pm$0.011 \nl  
4686 He \II   & 0.04 & 0.04 & $<$0.019\tablenotemark{bd} & $<$0.031\tablenotemark{bd}  &  $<$0.022\tablenotemark{bd} \nl  
4861 H$\beta$ & 0.00 & 0.00 & 1.000$\pm$0.006  & 1.000$\pm$0.010 & 1.000$\pm$0.010 \nl  
4959 [O \III] &-0.02 &-0.02 & 2.353$\pm$0.022  & 1.371$\pm$0.018 & 2.115$\pm$0.021 \nl  
5007 [O \III] &-0.03 &-0.03 & 6.896$\pm$0.090  & 4.150$\pm$0.070 & 5.925$\pm$0.079 \nl  
5876 He \I    &-0.23 &-0.23 & 0.115$\pm$0.013  & 0.144$\pm$0.018 & 0.111$\pm$0.013 \nl  
6300 [O \I]   &-0.30 &-0.30 & $<$0.030\tablenotemark{b}         & ---             & $<$0.018\tablenotemark{b} \nl
6312 [S \III] &-0.30 &-0.30 & $<$0.030\tablenotemark{b}         & ---             & $<$0.018\tablenotemark{b} \nl
6548 [N \II]  &-0.34 &-0.34 & ---              & ---             & ---             \nl  
6563 H$\alpha$&-0.34 &-0.34 & 2.826$\pm$0.356  & ---             & 2.849$\pm$0.359 \nl  
6584 [N \II]  &-0.34 &-0.34 & ---              & ---             & ---             \nl  
6678 He \I    &-0.35 &-0.35 & $<$0.036\tablenotemark{b}         & ---             & 0.055$\pm$0.012    \nl  
6717 [S \II]  &-0.36 &-0.36 & 0.095$\pm$0.015  & ---             & $<$0.028\tablenotemark{b} \nl  
6731 [S \II]  & -0.36 &-0.36 & 0.046$\pm$0.011  & ---             & $<$0.017\tablenotemark{b} \nl  
\nl
EW(Balmer abs) (\AA)           & &   & 0.0               &   2.0             &  2.0        \nl  
c(H$\beta$)                    & &   & 0.20$\pm$0.06     &   0.32$\pm$0.10   &  0.25$\pm$0.06 \nl  
F(H$\beta$) ($\times10^{14}$ erg s$^{-1}$cm$^{-2}$) & & &    1.24$\pm$0.03   &  0.91$\pm$0.04    & 0.75$\pm$0.03 \nl  
EW(H$\beta$)                   &  &  & 217$\pm$  2       &   60$\pm$  2      & 200$\pm$  2 \nl  
n$_e$ (cm$^{-3}$)              &  &  &   $<$100          &   ---             & --- \nl  
\enddata
\tablenotetext{a}{Extinction toward NGC 4861 and UM469 is due entirely to dust within
those systems, and thus we deredden UV lines using the SMC reddening law of Pr\'evot \etal\
1984, appropriate to their metallicity.
Toward T1345-420 we assume half of the extinction is due to Galactic forground
dust (Galactic curve) and the other half is intrinsic to T1345-420 (SMC type curve).  }
\tablenotetext{b}{Value indicates the 3$\sigma$ upper limit.  All other errors are 1$\sigma$.}
\tablenotetext{c}{[O~III] $\lambda\lambda$2321,2331 and the C~II] $\lambda$2326 multiplet
are blended at the resolution of the FOS.  The listed line strengths reflect the separation
estimate described in the text.}
\tablenotetext{d}{Includes only narrow, presumably nebular He~II. }
\end{planotable}

\begin{planotable}{lrrr}
\tablecaption{Calar Alto Dereddened Line Strengths}
\tablewidth{400pt}
\tablehead{
\colhead{Line} & 
\colhead{f($\lambda$)}   & 
\colhead{NGC 4861}   & 
\colhead{UM469}    \\ [.2ex]  
\colhead{  }   & 
\colhead{  }   & 
\colhead{${I_0}\over{I_{H\beta}}$ }   &
\colhead{${I_0}\over{I_{H\beta}}$ }    }
\startdata
3727 [O \II]  & 0.25 & 2.332$\pm$0.117  & 2.928$\pm$0.158     \nl  
3869 [Ne \III]& 0.22 & 0.500$\pm$0.030  & 0.419$\pm$0.042     \nl  
4101 H$\delta$& 0.18 & 0.238$\pm$0.015  & 0.388$\pm$0.031      \nl  
4343 H$\gamma$& 0.12 & 0.489$\pm$0.022  & 0.517$\pm$0.025      \nl  
4363 [O \III] & 0.11 & 0.084$\pm$0.009  & 0.045$\pm$0.013      \nl  
4471 He \I    & 0.09 & 0.061$\pm$0.008  & $<$0.039\tablenotemark{a}      \nl  
4686 He \II   & 0.04 & $<$0.015\tablenotemark{ab}  & $<$0.037\tablenotemark{ab}  \nl  
4861 H$\beta$ & 0.00 & 1.000$\pm$0.036  & 1.000$\pm$0.035  \nl  
4959 [O \III] &-0.02 & 1.517$\pm$0.052  & 1.255$\pm$0.043  \nl  
5007 [O \III] &-0.03 & 4.705$\pm$0.154  & 3.665$\pm$0.120  \nl  
5876 He \I    &-0.23 & 0.135$\pm$0.009  & 0.111$\pm$0.005  \nl  
6300 [O \I]   &-0.30 & 0.031$\pm$0.004  & 0.057$\pm$0.003  \nl
6312 [S \III] &-0.30 & 0.049$\pm$0.005  & 0.007$\pm$0.001  \nl
6548 [N \II]  &-0.34 & 0.027$\pm$0.004  & 0.079$\pm$0.005  \nl  
6563 H$\alpha$&-0.34 & 2.829$\pm$0.160  & 2.833$\pm$0.160  \nl  
6584 [N \II]  &-0.34 & 0.120$\pm$0.008  & 0.231$\pm$0.013  \nl  
6678 He \I    &-0.35 & 0.023$\pm$0.003  & 0.029$\pm$0.002 \nl  
6717 [S \II]  &-0.36 & 0.216$\pm$0.014  & 0.263$\pm$0.015 \nl  
6731 [S \II]  &-0.36 & 0.165$\pm$0.011  & 0.196$\pm$0.011 \nl  
7065 He I     &-0.40 & 0.022$\pm$0.004  & 0.029$\pm$0.002 \nl  
7136 [Ar \III]&-0.41 & 0.078$\pm$0.007  & 0.062$\pm$0.004 \nl  
7320 [O \II]  &-0.43 & 0.021$\pm$0.005  & 0.036$\pm$0.002 \nl  
7330 [O \II]  &-0.44 & 0.014$\pm$0.004  & 0.019$\pm$0.001  \nl  
9069 [S \III] &-0.59 & 0.137$\pm$0.021  & ---  \nl  
\nl
EW(Balmer abs) (\AA)           &   & 0.0               &   2.0            \nl  
c(H$\beta$)                    &   & 0.11$\pm$0.06     &   0.46$\pm$0.06  \nl  
F(H$\beta$) ($\times10^{14}$ erg s$^{-1}$cm$^{-2}$) & &    0.25$\pm$0.01   &  0.76$\pm$0.03   \nl  
EW(H$\beta$)                   &   & 190$\pm$  4       &   76$\pm$  3      \nl  
n$_e$ (cm$^{-3}$)              &   &   127$\pm$  88           &   95$\pm$  62     \nl  

\enddata
\tablenotetext{a}{Value indicates the 3$\sigma$ upper limit.}
\tablenotetext{b}{Includes only narrow, presumably nebular He~II. }
\end{planotable}

\clearpage

\begin{planotable}{lrr}
%\tablecaption{NGC 4861 Derived Parameters and Abundances   \label{elems1} }
\tablecaption{NGC 4861 Derived Parameters and Abundances  }
\tablewidth{380pt}
\tablehead{
\colhead{Parameter} & 
\colhead{HST FOS}   &
\colhead{Calar Alto}    } 
\startdata
T(O$^{++}$) (K)                     & 14210$\pm$ 440   & 14500$\pm$ 710  \nl  
T(O$^{+}$) (K)                      & 13300$\pm$ 670   & 13450$\pm$ 860  \nl  
O$^{+}$/H$^{+}$ ($\times10^{5}$)    & 1.53$\pm$0.22    & 2.88$\pm$0.36   \nl  
O$^{++}$/H$^{+}$ ($\times10^{5}$)   & 8.97$\pm$0.55    & 5.76$\pm$0.56  \nl  
O$^{+++}$/H$^{+}$ ($\times10^{5}$)  & $<$0.18          & $<$0.10  \nl  
O/H ($\times10^{5}$)                & 10.69$\pm$0.59    & 8.73$\pm$0.67  \nl  
12+log(O/H)                         & 8.03$^{+0.02}_{-0.02}$   & 7.94$^{+0.03}_{-0.03}$  \nl  
                                    &                  &                 \nl 
S$^+$/H$^+$ ($\times10^{7}$)        & 1.74$\pm$0.48    & 4.61$\pm$0.52  \nl 
S$^{++}$/H$^+$ ($\times10^{7}$)     & $<$19.2          & 29.25$\pm$4.45 \nl 
S$^{3+}$/H$^+$ ($\times10^{7}$)     & ---              & 6.83$\pm$1.20  \nl 
S/H              ($\times10^{7}$)   & ---              & 40.68$\pm$5.57 \nl 
log(S/O)                            & ---              & -1.33$^{+0.06}_{-0.07}$ \nl 
                                    &                  &                         \nl 
N$^+$/O$^+$ (${\lambda6584}/{\lambda3727}$)                        & ---              & 0.040$\pm$0.006           \nl  
log(N/O)(${\lambda6584}/{\lambda3727}$)                            & ---              & -1.40$^{+0.06}_{-0.07}$             \nl  
log(N/O)(${\lambda6584}/{\lambda7325}$)                           & ---              & -1.52$^{+0.09}_{-0.11}$             \nl  
$<$log(N/O)$>$                        		        	  & ---              & -1.44$^{+0.05}_{-0.06}$             \nl  
                                    &                  &                                     \nl 
C$^{+}$/O$^{+}$({$\lambda2326/\lambda3727$})              & 0.307$\pm$0.067   & ---            \nl  
log(C$^{+}$/O$^{+}$) ({$\lambda2326/\lambda3727$})        & -0.512$^{+0.086}_{-0.107}$  & ---            \nl  
log(C/O) = log(C$^{+}$/O$^{+}$)+ICF		   	  & -0.435$^{+0.086}_{-0.107}$  & ---            \nl  
C$^{++}$/O$^{++}$({$\lambda1909/\lambda5007$})& 0.308$^{+0.079}_{-0.064}$   & ---            \nl  
log(C/O) ({$\lambda1909/\lambda5007$})        & -0.511$^{+0.100}_{-0.101}$  & ---            \nl  
C$^{++}$/O$^{++}$({$\lambda1909/\lambda1661$})& $>$0.018                    & ---            \nl  
log(C$^{++}$/O$^{++}$) ({$\lambda1909/\lambda1661$})        & $>$-1.74                    & ---            \nl  
                                    &                                       &                \nl 
Ne$^{++}$/O$^{++}$                  & 0.165$\pm$0.020          & 0.196$\pm$0.028         \nl  
log(Ne/O)                           & -0.78$^{+0.05}_{-0.06}$  & -0.71$^{+0.06}_{-0.07}$ \nl  
 & \nl
Si$^{++}$/C$^{++}$                  & 0.094$\pm$0.019        & ---         \nl
Si/C                  		    & 0.134$\pm$0.026        & ---         \nl
log(Si/O)                           & -1.38$\pm$0.13         & ---         \nl
                                    &                        &                       \nl 
He$^+$/H$^+$ ($\lambda$4471)        & 0.087$\pm$0.014          & 0.132$\pm$0.019   \nl  
He$^+$/H$^+$ ($\lambda$5876)        & 0.094$\pm$0.012          & 0.112$\pm$0.010  \nl  
He$^+$/H$^+$ ($\lambda$6678)        & 0.106$\pm$0.020          & 0.068$\pm$0.011   \nl  
\enddata
\end{planotable}

\begin{planotable}{lrrr}
%\tablecaption{UM469 Derived Parameters and Abundances   \label{elems1} }
\tablecaption{UM469 Derived Parameters and Abundances  }
\tablewidth{380pt}
\tablehead{
\colhead{Parameter} & 
\colhead{HST FOS}   &
\colhead{Calar Alto}    } 
\startdata
T(O$^{++}$) (K)                     & 12840$\pm$ 840   & 12390$\pm$1000  \nl  
T(O$^{+}$) (K)                      & 12667$\pm$ 980   & 12440$\pm$1000  \nl  
O$^{+}$/H$^{+}$ ($\times10^{5}$)    & 2.04$\pm$0.38    & 4.73$\pm$0.97   \nl  
O$^{++}$/H$^{+}$ ($\times10^{5}$)   & 7.13$\pm$0.99    & 6.97$\pm$1.76  \nl  
O$^{+++}$/H$^{+}$ ($\times10^{5}$)  & $<$0.21          & $<$0.41  \nl  
O/H ($\times10^{5}$)                & 9.37$\pm$1.06    & 12.11$\pm$2.01  \nl  
12+log(O/H)                         & 7.97$^{+0.05}_{-0.05}$   & 8.08$^{+0.07}_{-0.08}$  \nl  
                                    &                  &                 \nl 
S$^+$/H$^+$ ($\times10^{7}$)        & ---              & 6.51$\pm$0.92   \nl 
S$^{++}$/H$^+$ ($\times10^{7}$)     & ---              & 6.72$\pm$2.24 \nl 
S$^{3+}$/H$^+$ ($\times10^{7}$)     & ---              & 1.94$\pm$0.72 \nl 
S/H              ($\times10^{7}$)   & ---              & 15.18$\pm$2.90\nl 
log(S/O)                            & ---              & -1.90$^{+0.10}_{-0.13}$ \nl 
                                    &                  &                 \nl 
N$^+$/O$^+$ (${\lambda6584}/{\lambda3727}$)                       & ---              & 0.068$\pm$0.015       \nl  
log(N/O)(${\lambda6584}/{\lambda3727}$)                          & ---              & -1.26$^{+0.09}_{-0.11}$  \nl  
log(N/O)(${\lambda6584}/{\lambda7325}$)                          & ---              & -1.29$^{+0.12}_{-0.16}$  \nl  
$<$log(N/O)$>$                         				 & ---              & -1.27$^{+0.08}_{-0.10}$  \nl  
                                    &                  &                 \nl 
C$^{+}$/O$^{+}$({$\lambda2326/\lambda3727$})              & 0.265$\pm$0.074   & ---            \nl  
log(C$^{+}$/O$^{+}$) ({$\lambda2326/\lambda3727$})        & -0.576$^{+0.089}_{-0.113}$  & ---            \nl  
log(C/O) = log(C$^{+}$/O$^{+}$)+ICF			  & -0.498$^{+0.089}_{-0.113}$  & ---            \nl 
C$^{++}$/O$^{++}$({$\lambda1909/\lambda5007$})& 0.427$^{+0.257}_{-0.171}$  &  ---         \nl  
log(C/O) ({$\lambda1909/\lambda5007$})        & -0.369$^{+0.204}_{-0.222}$ & ---          \nl  
C$^{++}$/O$^{++}$({$\lambda1909/\lambda1661$})& $>$0.47                    & ---          \nl  
log(C$^{++}$/O$^{++}$) ({$\lambda1909/\lambda1661$})        & $>$-1.32                   & ---          \nl  
                                    &                        &                         \nl 
Ne$^{++}$/O$^{++}$                  &  0.149$\pm$0.034       & 0.226$\pm$0.084         \nl  
log(Ne/O)                           & -0.83$^{+0.09}_{-0.11}$& -0.65$^{+0.14}_{-0.20}$ \nl  

Si$^{++}$/O$^{++}$                  & $<$0.277	         & ---                     \nl
Si/C	 	                    & $<$0.395           & ---                     \nl
log(Si/O)                           & $<$-0.772          & ---                     \nl
                                    &                        &                         \nl 
He$^+$/H$^+$ ($\lambda$4471)        & 0.112$\pm$0.022          & 0.087$\pm$0.029  \nl  
He$^+$/H$^+$ ($\lambda$5876)        & 0.116$\pm$0.018          & 0.089$\pm$0.016  \nl  
He$^+$/H$^+$ ($\lambda$6678)        & ---                      & 0.085$\pm$0.015  \nl  
\enddata
\end{planotable}

\begin{planotable}{lrrr}
%\tablecaption{T1345-420 Derived Parameters and Abundances   \label{elems1} }
\tablecaption{T1345-420 Derived Parameters and Abundances  }
\tablewidth{380pt}
\tablehead{
\colhead{Parameter} & 
\colhead{HST FOS} &  
\colhead{CTM86}   }
\startdata
T(O$^{++}$) (K)                     & 12520$\pm$ 740  & 13320$\pm$ 720  \nl  
T(O$^{+}$) (K)                      & 12509$\pm$ 900  & 12896$\pm$ 880  \nl  
O$^{+}$/H$^{+}$ ($\times10^{5}$)    &  1.66$\pm$0.28  & 2.35$\pm$0.37  \nl  
O$^{++}$/H$^{+}$ ($\times10^{5}$)   & 10.94$\pm$1.42  & 8.20$\pm$0.97  \nl  
O$^{+++}$/H$^{+}$ ($\times10^{5}$)  & 0.26$\pm$0.12   & 0.24$\pm$0.05  \nl  
O/H ($\times10^{5}$)                & 12.85$\pm$1.45  & 10.79$\pm$1.04  \nl  
12+log(O/H)                         & 8.11$^{+0.05}_{-0.05}$ & 8.03$^{+0.04}_{-0.04}$ \nl  
                                    &                 &   \nl 
S$^+$/H$^+$ ($\times10^{7}$)        & $<$0.63         & 3.45$\pm$0.78  \nl 
S$^{++}$/H$^+$ ($\times10^{7}$)     & $<$17.5         & ---  \nl 
S$^{3+}$/H$^+$ ($\times10^{7}$)     & ---             & ---  \nl 
S/H              ($\times10^{7}$)   & ---             & ---  \nl 
log(S/O)                            & ---             & ---  \nl 
                                    &                 &   \nl 
N$^+$/O$^+$ (${\lambda6584}/{\lambda3727}$)                         & ---             & 0.02$\pm$0.01  \nl  
log(N/O)(${\lambda6584}/{\lambda3727}$)                            & ---             & -1.62$^{+0.09}_{-0.12}$  \nl  
                                    &                 &   \nl 
C$^{+}$/O$^{+}$({$\lambda2326/\lambda3727$})& 0.170$\pm$0.046   & ---            \nl  
log(C$^{+}$/O$^{+}$) ({$\lambda2326/\lambda3727$})        & -0.769$^{+0.103}_{-0.136}$  & ---            \nl  
log(C/O) = log(C$^{+}$/O$^{+}$)+ICF             	  & -0.690$^{+0.103}_{-0.136}$  & ---            \nl  
C$^{++}$/O$^{++}$({$\lambda1909/\lambda5007$})& 0.196$^{+0.059}_{-0.049}$  & --- \nl  
log(C/O) ({$\lambda1909/\lambda5007$})        & -0.707$^{+0.122}_{-0.165}$   & --- \nl  
C$^{++}$/O$^{++}$({$\lambda1909/\lambda1661$})& $>$0.043                   & --- \nl  
log(C$^{++}$/O$^{++}$) ({$\lambda1909/\lambda1661$})        & $>$-1.36                   & --- \nl  
                                    &                                      & \nl 
Ne$^{++}$/O$^{++}$                  & 0.167$\pm$0.034                      & 0.173$\pm$0.033\nl  
log(Ne/O)                           & -0.78$^{+0.08}_{-0.10}$              & -0.76$^{+0.08}_{-0.09}$ \nl  
			            &                                      & \nl 
Si$^{++}$/O$^{++}$                  & $<$0.257                             & --- \nl
Si/C		                    & $<$0.367                             & --- \nl
log(Si/O)                           & $<$-1.14                             & --- \nl
                                    &                                      &      \nl 
He$^+$/H$^+$ ($\lambda$4471)        & 0.103$\pm$0.025 & 0.046$\pm$0.012          \nl  
He$^+$/H$^+$ ($\lambda$5876)        & 0.089$\pm$0.013 & 0.081$\pm$0.011           \nl  
He$^+$/H$^+$ ($\lambda$6678)        & 0.157$\pm$0.036 & 0.119$\pm$0.025           \nl  
\enddata
\end{planotable}